\documentclass[%
 reprint,
 amsmath,amssymb,
 aps,
]{revtex4-2}

\usepackage{amsmath}

\usepackage{graphicx}
\usepackage{dcolumn}
\usepackage{bm}

\usepackage{float}

\begin{document}

\preprint{APS/123-QED}

\title{Teukolsky Master Equation for a Kerr-like metric with Mass Quadrupole Moment}

\author{Pedro G\'omez}%
\email{pedro.gomezovares@ucr.ac.cr}
\affiliation{School of Physics \\ 
University of Costa Rica}

\author{Francisco Frutos-Alfaro}%
\email{francisco.frutos@ucr.ac.cr}
\homepage{http://www.cinespa.ucr.ac.cr/}
\affiliation{Space Research Center \\ School of Physics \\ 
University of Costa Rica}

\author{Adri\'an Eduarte-Rojas}
\email{adrian.eduarte@ucr.ac.cr}
\affiliation{School of Physics \\ 
University of Costa Rica}

\author{Antonio Banichevich}
\email{antonio.banichevich@ucr.ac.cr}
\affiliation{Space Research Center \\ School of Physics \\ 
University of Costa Rica}




\date{\today}

\begin{abstract}

The Teukolsky Master Equation (TME) is a powerful technique used to study perturbations of a spacetime with gravitational, electromagnetic, and neutrino components. The well-known Kerr metric is a suitable candidate for this formalism because it is a Petrov type D metric. However, the Kerr metric represents the exterior spacetime of a perfect sphere and only black holes are described using this metric. For another astrophysical object, such as fast-spinning neutron stars, the Kerr metric is not valid due to the rotation induced deformations to the gravitational source. The main goal of the present work is to show a TME for a Kerr-like metric (KL) with a mass quadrupole moment parameter $q$. This new parameter will represent deformations from a perfect sphere; therefore, KL is a better candidate to study perturbations surrounding spinning stars. 

\end{abstract}


\maketitle


\section{\label{sec:Intro}Introduction}

It is often find that many physical systems can be described as a known idealized system with some variation. If these variations are sufficiently small, one way to study them is through perturbation theory, which has been regularly used to analyze various phenomena in different branches of physics, such as quantum mechanics or even general relativity.  The last one describes the gravitational interaction with unprecedented accuracy, allowing us to understand the field produced by stars, planets, or compact objects, among other applications.

For the case of compact objects modeled as perfect rotating spheres, whose exterior gravitational field is described by the Kerr metric \cite{kerr1963gravitational}. Perturbations around the object were studied by Saul Teukolsky in 1973 \cite{teukolsky1973perturbations}, using a particular formalism of general relativity known as the Newman--Penrose (NP) formalism. He derived a single equation (\ref{EQ:MASTERKERR}) describing gravitational, electromagnetic, and scalar perturbations, known as the Teukolsky \textit{Master} equation (TME). Likewise, the perturbation formalism developed by Teukolsky is valid for a family of metrics known as Petrov type D.

\begin{eqnarray}\label{EQ:MASTERKERR}
 & &\left[ \dfrac{(r^{2}+a^{2})^{2}}{\Delta}-a^{2}\sin^{2}\theta\right]\dfrac{\partial^{2}\psi}{\partial t^{2}}+\dfrac{4Mar}{\Delta}\dfrac{\partial^{2}\psi}{\partial t\partial\varphi}  \nonumber \\
 &+&\left[\dfrac{a^{2}}{\Delta}-\dfrac{1}{\sin^{2}\theta}\right]\dfrac{\partial^{2}\psi}{\partial \varphi^{2}} -\Delta^{-s}\dfrac{\partial}{\partial r}\left(\Delta^{s+1}\dfrac{\partial\psi}{\partial r}\right) \nonumber \\
 &-&\dfrac{1}{\sin\theta}\dfrac{\partial}{\partial\theta}\left(\sin\theta\dfrac{\partial\psi}{\partial \theta}\right)-2s\left[\dfrac{a(r-M)}{\Delta}+\dfrac{i\cos\theta}{\sin^{2}\theta}\right]\dfrac{\partial\psi}{\partial\varphi} \nonumber \\
 &-&2s\left[\dfrac{M(r^{2}-a^{2})}{\Delta}-r-ia\cos\theta \right]\dfrac{\partial\psi}{\partial t} \nonumber \\
 &+&(s^{2}\cot^{2}\theta-s)\psi=4\pi\Sigma T ,
\end{eqnarray}

\noindent where $M$, $a$, $\Delta$, and $\Sigma$ have their usual meanings for the Kerr metric, $s$ is the \textit{spin weight}, and $\psi$ corresponds to any of the perturbative fields with their respective source term $T$, which take the values indicated in Table \ref{tabla:teu} \cite{teukolsky1973perturbations}. The equation can also be used to describe perturbations in Schwarzschild spacetime by setting $a=0$, and in Minkowski spacetime in spherical coordinates with $a=0$ and $M=0$.

\begin{table}[t]
    \centering
    \renewcommand{\arraystretch}{1.45}
    \setlength{\tabcolsep}{14pt} 
    \large 
    \begin{tabular}{c c c}
        \hline
        $\boldsymbol{\psi}$ & $\boldsymbol{s}$ & $\boldsymbol{T}$ \\
        \hline
        $\Phi^{(1)}$
        & $0$
        & $T^{(1)}$ \\
        \hline
        $\phi_{0}^{(1)}$
        & $1$
        & $J_{0}^{(1)}$ \\[2pt]
        
        $\rho^{-2}\phi_{2}^{(1)}$
        & $-1$
        & $\rho^{-2}J_{2}^{(1)}$ \\
        \hline
        $\Psi_{0}^{(1)}$
        & $2$
        & $2T_{0}^{(1)}$ \\[2pt]
        
        $\rho^{-4}\Psi_{4}^{(1)}$
        & $-2$
        & $2\rho^{-4}T_{4}^{(1)}$ \\
        \hline
    \end{tabular}
    \caption{$\psi$, $s$ and $T$ for the Teukolsky master equation.}
    \label{tabla:teu}
\end{table}

It is possible to demonstrate that the fields $\Psi_{0}^{(1)}$, $\Psi_{4}^{(1)}$, $\phi_{0}^{(1)}$, and $\phi_{2}^{(1)}$ contain all the non-trivial information of the complete perturbative system \cite{fackerell1972weak,wald1973perturbations}, and once the solutions to the Teukolsky master equation are obtained, the solutions for the entire coupled perturbation system can be constructed from them \cite{wald1978construction}, so it is not necessary to derive equations for any of the other Newman--Penrose quantities.

However, compact objects in general are not perfect spheres and may present deformations. One way to introduce this deformation into the model is by adding a quadrupole parameter to the object, so it becomes a spheroid rather than a perfect sphere. For an object with these characteristics, the exterior gravitational field can be described by various metrics belonging to different Petrov families, among which is the KL (\textit{Kerr-Like}) metric \cite{frutos2016approximate}, which was derived using Fodor--Hoenselaers--Perj\'es algorithm construct the $n$-th multipole moment of an axisymmetric, stationary, rotating spacetime by using the Ernst potential expanded in Taylor series and the Kerr metric as a seed \cite{Carmeli, fodor, frutos2016approximate}. This work focuses on deriving a TME for gravitational, electromagnetic, and scalar perturbations using the Newman--Penrose formalism, employing the KL solution as the base metric, with the aim that it can be used for astrophysical applications in the future. There are plenty of works involving the NP formalism and other advanced techniques such as Geroch--Held--Penrose (GHP) formalism which take advantage if the spacetime possesses one or two preferred null directions and the spherical symmetry \cite{Durkee_2010, BargenoGHP, geroch1973space,  SmarrEq}.

Despite the apparent simplicity of Einstein’s field equations, they correspond to a system of coupled nonlinear differential equations, so their solution is significantly complex, even in the simplest cases. Furthermore, in general, the components of the energy-momentum tensor may contain metric terms, so to find the full solution it is necessary to solve simultaneously the equations for the spacetime metric and for the distribution of matter and non-gravitational fields \cite{wald2010general}.

If we know the equations governing the behavior of some non-gravitational field or matter distribution that are valid in the context of special relativity, we can use minimal coupling to extend these equations so that they are also valid in general relativity.

In the case of a scalar field $\Phi$ with mass $m$, it will satisfy the \textit{Klein-Gordon equation} (in the so-called form minimally coupled case), substituting partial derivatives with covariant derivatives, and the Minkowski metric by the metric of a curved spacetime \cite{wald2010general, birrell}

\begin{align}\label{KGEQ}
    \nabla^{\mu}\nabla_{\mu} \Phi - m^{2} \Phi = 0.
\end{align}

Its energy-momentum tensor is given by:

\begin{align}\label{eq:Tklein}
    T_{\mu\nu} = \nabla_{\mu} \Phi \nabla_{\nu} \Phi - \dfrac{1}{2} g_{\mu\nu} \left( \nabla_{\alpha} \Phi \nabla^{\alpha} \Phi + m^{2} \Phi^{2} \right).
\end{align}

Performing the same procedure for the electromagnetic field, we obtain the \textit{Maxwell equations} in curved spacetime \cite{misner2017gravitation}

\begin{eqnarray}\label{MQE}
\nabla_{\nu} F^{\mu\nu} = 4\pi J^{\mu}, \nonumber \\
\nabla_{\alpha} F_{\mu\nu} + \nabla_{\mu} F_{\nu\alpha} + \nabla_{\nu} F_{\alpha\mu} = 0,
\end{eqnarray}

where $J^{\mu}$ is the \textit{four-current density vector}, and $F^{\mu\nu}$ is the \textit{electromagnetic field tensor}, which is \textit{antisymmetric}: $F^{\mu\nu} = -F^{\nu\mu}$.

We can also find the energy-momentum tensor for the electromagnetic field \cite{wald2010general}

\begin{align}\label{eq:Tmax}
    T_{\mu\nu} = \dfrac{1}{4\pi} \left( F_{\mu\alpha} F_{\nu}^{\ \alpha} - \dfrac{1}{4} g_{\mu\nu} F_{\alpha\beta} F^{\alpha\beta} \right).
\end{align}

Eq.~(\ref{MQE}) together with the Einstein field equation using the electromagnetic energy-momentum tensor is also known as the \textit{Einstein-Maxwell equations}.

We can use the above equations, along with Einstein’s field equations, to describe the coupled dynamics of electromagnetic or scalar fields with the curvature of spacetime. However, due to the difficulty of simultaneously solving these equations, in some cases \cite{PhysRevLett.109.081102,PhysRevLett.27.529,teukolsky1973perturbations} it is relevant to work with \textit{test fields}, where only the dynamics of the fields on a fixed background metric spacetime are studied. This approximation will be used for the derivation of electromagnetic and scalar perturbations.

This paper is organized as follows. Section \ref{Sec:Metric} introduces the spacetime metric considered throughout this work. Section \ref{ch:FKLNP} presents the development of the Newman--Penrose formalism applied to this metric. Section \ref{ch:PFKL} introduces the perturbation technique using the NP formalism. Finally, section \ref{Sec:Conclusions} summarizes our main conclusions.

\section{Kerr-like Metric with Mass Quadrupole Moment} \label{Sec:Metric}

The KL metric is an approximate solution of the Einstein Field Equations that represents the exterior gravitational field of a compact object with quadrupolar deformation in rotation found by Francisco Frutos \cite{frutos2016approximate}. The solution has the following form

\begin{eqnarray}\label{Eq:KL}
    ds^{2} &=& g_{tt} dt^{2} + g_{rr} dr^{2} + g_{\theta\theta} d\theta^{2} + g_{\varphi\varphi} d\varphi^{2} \nonumber \\
    &+& 2 g_{t\varphi} dt d\varphi,
\end{eqnarray}

\noindent with

\begin{eqnarray}\label{KL}
    g_{tt} &=& \frac{e^{-2\psi}}{\Sigma} \left( a^{2} \sin^{2}\theta - \Delta \right),   \nonumber \\
    g_{t\varphi} &=& -\frac{2 M a r}{\Sigma} \sin^{2} \theta,  \nonumber \\
    g_{rr} &=& \Sigma \frac{e^{2\chi}}{\Delta}, \\
    g_{\theta\theta} &=& \Sigma e^{2\chi}, \nonumber \\
    g_{\varphi\varphi} &=& \frac{e^{2\psi}}{\Sigma} \left[ (r^{2} + a^{2})^{2} - a^{2} \Delta \sin^{2} \theta \right] \sin^{2} \theta . \nonumber
\end{eqnarray}

\noindent where $\Sigma = r^{2} + a^{2} \cos^{2} \theta$, $\Delta = r^{2} - 2 M r + a^{2}$, and the functions $\psi$ and $\chi$ are defined as

\begin{eqnarray}\label{eq:FKL CUAD}
    \psi &=& \frac{q}{r^{3}} P_{2} + 3 \frac{q M}{r^{4}} P_{2}, \nonumber \\
    \chi &=& \frac{q}{r^{3}} P_{2} + \frac{q M}{r^{4}} \left( -\frac{1}{3} + \frac{5}{3} P_{2} + \frac{5}{3} P_{2}^{2} \right) \\
    &+& \frac{q^{2}}{r^{6}} \left( \frac{2}{9} - \frac{2}{3} P_{2} - \frac{7}{3} P_{2}^{2} + \frac{25}{9} P_{2}^{3} \right).  \nonumber
\end{eqnarray}

The function $P_{2}$ is used as an abbreviation for $P_{2}(\cos\theta) = (3 \cos^{2} \theta - 1)/2$, which is the second Legendre polynomial with argument $\cos\theta$, $M$ is the mass of the object, $J$ is the angular momentum, $a = J / M$ is the rotation parameter, and $q$ is the quadrupole parameter. This solution is valid up to order $O(a q^{2}, a^{2} q, M q^{2}, M^{2} q, M a q, q^{3})$, is stationary, and has axial symmetry, so the rotation axis coincides with the object's symmetry axis.

The KL metric has the same form as the Kerr metric in Boyer-Lyndquist coordinates, except for the presence of the functions $\psi$ and $\chi$, which vanish in the limiting case $q=0$, so it is clearly seen that the KL metric is an extension of the Kerr metric accounting for the object's deformation. Also, as with the Kerr metric, performing the substitutions $\Delta \rightarrow \Delta + e^{2}$, and $2 M r \rightarrow r^{2} + a^{2} - \Delta$ in the $g_{t\varphi}$ coefficient, a solution including the object's charge $e$ is found, thus extending the Kerr-Newman solution \cite{frutos2016approximate2}.

The KL solution has other important limiting cases. In the case $a=0$ and redefining $q \rightarrow 2 q M^{3} / 15$, the metric reduces to the expansion up to second order in $q$ of the Erez--Rosen metric \cite{ERvsHT}. If the KL metric is expanded to first order in $q$, second order in $a$ and $M$, and the substitution $q \rightarrow J^{2} / M - Q$ is made, it reduces to the Hartle--Thorne metric expansion to the same order \cite{frutos2016approximate}.

If we expand the metric to linear order in $q$, and take $a=0$, the component $g_{tt}$ takes the following form

\begin{eqnarray} 
    g_{tt}  &\approx& - \left( 1 - \frac{2 M}{r} - \frac{2 q P_{2}}{r^{3}} \right).
\end{eqnarray}

Using the Newtonian limit $(g_{00} = g_{tt} = -(1 + 2 \Phi))$, we find that the gravitational potential $\Phi$ is given by

\begin{align}
\Phi = - \frac{M}{r} - \frac{q P_{2}}{r^{3}},
\end{align}

which corresponds to the Newtonian gravitational potential, in geometrized units, for an oblate spheroid \cite{capderou2014handbook}.

The form of the KL solution, and its similarity to the Kerr metric, will allow calculations and derivations that can be directly compared analytically with results found for the Kerr solution. This is the motivation for using the KL metric to study perturbations around rotating compact objects with small deformations. Moreover, there was developed numerical solutions of the KL to study the chaotic behavior outside the gravitational source \cite{adrian}.

The KL metric will be used in its linear limit \cite{frutos2014perturbation}, in which the metric components have the same form as Eq.~(\ref{KL}), but the functions $\psi$ and $\chi$ reduce to

\begin{eqnarray}\label{CHI}
\psi = \chi = \frac{q}{r^{3}} P_{2}.
\end{eqnarray}

This limit neglects terms of order $O(q M, q a, q^{2})$, and its use will be justified in section \ref{ch:FKLNP}.

\section{\label{ch:FKLNP}  The KL Solution in the Newman-Penrose Formalism}

The Teukolsky formalism has been extended to include both type D solutions of the Einstein field equations that incorporate electromagnetic fields in the background metric \cite{bose1975studies}, as well as fully arbitrary type D solutions depending on a large number of parameters \cite{dudley1979covariant,ahmed1986perturbations}. However, the presence of background electromagnetic fields, as well as other energy sources in these arbitrary solutions, leads to coupling of the equations for the perturbative fields. Therefore, in order to obtain decoupled equations, it is necessary to make approximations on the different parameters on which the solutions depend. A complete description of the NP formalism is shown in the appendix.

Likewise, for vacuum solutions, the fact that we can find decoupled equations for the quantities $\Psi_{0}^{(1)}$, $\Psi_{4}^{(1)}$, $\phi_{0}^{(1)}$, and $\phi_{2}^{(1)}$ is only possible if the metric is of type D \cite{stewart1974perturbations}, so for more general solutions we cannot derive an exact master equation that describes the perturbations.

Before performing perturbations of the KL metric in the NP formalism, we need to first describe this solution within the formalism itself. This description has not been carried out previously, so in this section, we derive all the NP quantities for the KL solution. Additionally, we study the Petrov type of this metric using this formalism.

\subsection{Null Tetrad for the KL Metric}

To analyze the KL metric within the Newman-Penrose formalism, we must first construct a null tetrad from which the remaining quantities will be built. This tetrad is not unique; however, in order to make a direct comparison with the NP expressions for the Kerr metric, we seek a tetrad that has a similar form to the Kinnersley tetrad \cite{kinnertetrad, kinnertetrad2prd}.

The only difference between the form of the Kerr metric and the KL metric lies in exponential factors, so we can write the KL solution as

\begin{eqnarray}\label{eq:FKLkerr expo}
    g_{tt} &=& e^{-2\psi}g_{tt}^{K} , \nonumber \\
    g_{rr} &=& e^{2\chi}g_{rr}^{K} , \nonumber \\ 
    g_{\theta\theta} &=& e^{2\chi}g_{\theta\theta}^{K}, \\
    g_{\varphi\varphi} &=& e^{2\psi}g_{\varphi\varphi}^{K}  , \nonumber \\ 
    g_{t\varphi} &=& g_{t\varphi}^{K}. \nonumber
\end{eqnarray}

where the variables with index $K$ in this chapter correspond to the expressions for the Kerr metric in Boyer-Lyndquist coordinates. From Eq.~(\ref{eq:metricatetradas}), we can express the components of both metrics in terms of their respective null tetrads, so we obtain

\begin{eqnarray}
   -2l_{t}n_{t}+2m_{t}\bar{m}_{t} &=& e^{-2\psi}(-2l_{t}^{K}n_{t}^{K}  +2m_{t}^{K}\bar{m}_{t}^{K}) ,\nonumber \\ 
   -2l_{r}n_{r}+2m_{r}\bar{m}_{r} &=& e^{2\chi}(-2l_{r}^{K}n_{r}^{K}+2m_{r}^{K}\bar{m}_{r}^{K}) ,\nonumber \\  
   -2l_{\theta}n_{\theta}+2m_{\theta}\bar{m}_{\theta} &=& e^{2\chi}(-2l_{\theta}^{K}n_{\theta}^{K}+2m_{\theta}^{K}\bar{m}_{\theta}^{K}) , \\
   -2l_{\varphi}n_{\varphi}+2m_{\varphi}\bar{m}_{\varphi} &=& e^{2\psi}(-2l_{\varphi}^{K}n_{\varphi}^{K}+2m_{\varphi}^{K}\bar{m}_{\varphi}^{K}) ,\nonumber \\
   -l_{t}n_{\varphi} - n_{t}l_{\varphi} &+& m_{t}\bar{m}_{\varphi}+\bar{m}_{t}m_{\varphi} = -l_{t}^{K}n_{\varphi}^{K} \nonumber \\
   &-& n_{t}^{K}l_{\varphi}^{K} + m_{t}^{K}\bar{m}_{\varphi}^{K} + \bar{m}_{t}^{K}m_{\varphi}^{K}\ . \nonumber
\end{eqnarray}

It is easy to see that the previous relations hold if we propose that the null tetrad of the KL metric takes the following form

\begin{eqnarray}\label{eq tetradkerrlikenot}
    k_{t} &=& e^{-\psi}k_{t}^{K} ,  \nonumber \\
    k_{r}  &=& e^{\chi}k_{r}^{K} , \\
    k_{\theta} &=& e^{\chi}k_{\theta}^{K} ,   \nonumber  \\
    k_{\varphi} &=& e^{\psi}k_{\varphi}^{K} .  \nonumber
\end{eqnarray}

\noindent where $k_{\mu}$ represents any vector in the null tetrad.

We now proceed to verify that the proposed tetrad satisfies the normalization conditions Eq.~(\ref{eq:normalizatetrada}). To do this, we use the fact that the inverse KL metric can be written in terms of the inverse Kerr metric as

\begin{eqnarray}
    g^{tt} &=& e^{2\psi}g^{tt}_{K} , \nonumber \\
    g^{rr} &=& e^{-2\chi}g^{rr}_{K} , \nonumber \\
    g^{\theta\theta} &=& e^{-2\chi}g^{\theta\theta}_{K} ,  \\
    g^{\varphi\varphi} &=& e^{-2\psi}g^{\varphi\varphi}_{K}, \nonumber \\
    g^{t\varphi} &=& g^{t\varphi}_{K} . \nonumber 
\end{eqnarray}

\noindent so that when calculating the inner product between two arbitrary tetrad vectors, the exponential factors cancel out and we obtain

\begin{eqnarray}
    k^{\mu}v_{\mu} &=& g^{\mu\nu}k_{\nu}v_{\mu}  \nonumber \\
    &=& g^{tt}k_{t}v_{t} + g^{rr}k_{r}v_{r} + g^{\theta\theta}k_{\theta}v_{\theta} \nonumber  \\
    &+& g^{\varphi\varphi}k_{\varphi}v_{\varphi} + g^{t\varphi}k_{\varphi}v_{t} + g^{\varphi t}k_{t}v_{\varphi} \nonumber \\
    &=& g^{tt}_{K}k_{t}^{K}v_{t}^{K} + g^{rr}_{K}k_{r}^{K}v_{r}^{K} + g^{\theta\theta}_{K}k_{\theta}^{K}v_{\theta}^{K} \\
    &+& g^{\varphi\varphi}_{K}k_{\varphi}^{K}v_{\varphi}^{K} + g^{t\varphi}_{K}k_{\varphi}^{K}v_{t}^{K} + g^{\varphi t}_{K}k_{t}^{K}v_{\varphi}^{K} \nonumber \\
    &=& g^{\mu\nu}_{K}k_{\nu}^{K}v_{\mu}^{K} \nonumber  \\
    &=& k^{\mu}_{K}v_{\mu}^{K}  . \nonumber 
\end{eqnarray}

We thus observe that the inner products between the null tetrad vectors are the same for the KL and Kerr metrics, confirming that the tetrad defined in Eq.~(\ref{eq tetradkerrlikenot}) is indeed a valid null tetrad for the KL solution.

We can also express the contravariant components of the null tetrad of the KL metric in terms of those of the Kerr metric as

\begin{eqnarray}
    k^{t} &=& g^{t\nu}k_{\nu} = g^{tt}k_{t} + g^{t\varphi}k_{\varphi} \nonumber \\
    &=& e^{\psi}(g^{tt}_{K}k_{t}^{K} + g^{t\varphi}_{K}k_{\varphi}^{K}) = e^{\psi}k^{t}_{K}\ , \nonumber  \\
    k^{r} &=& g^{r\nu}k_{\nu} = g^{rr}k_{r} \nonumber \\
    &=& e^{-\chi}g^{rr}_{K}k_{r}^{K} = e^{-\chi}k^{r}_{K}\ , \nonumber\\
    k^{\theta} &=& g^{\theta\nu}k_{\nu} = g^{\theta\theta}k_{\theta}  \\
    &=& e^{-\chi}g^{\theta\theta}_{K}k_{\theta}^{K} = e^{-\chi}k^{\theta}_{K}\ , \nonumber \\
    k^{\varphi} &=& g^{\varphi\nu}k_{\nu} = g^{\varphi\varphi}k_{\varphi} + g^{\varphi t}k_{t} \nonumber \\
    &=& e^{\psi}(g^{\varphi\varphi}_{K}k_{\varphi}^{K}+g^{\varphi t}_{K}k_{t}^{K}) = e^{\psi}k^{\varphi}_{K} . \nonumber
\end{eqnarray}

\noindent therefore, using the Kinnersley tetrads \cite{kinnertetrad, kinnertetrad2prd} we can explicitly write the components of the null tetrad for the KL solution

\begin{align}\label{tetradaFKLnula}
\begin{aligned}
   & l^{\mu}=\left[\dfrac{e^{\psi}(r^{2}+a^{2})}{\Delta},e^{-\chi},0,\dfrac{e^{-\psi}a}{\Delta}\right]  ,\\
   & n^{\mu}=\left[e^{\psi}(r^{2}+a^{2}),-e^{-\chi}\Delta,0,e^{-\psi}a\right]/2\Sigma ,\\
   & m^{\mu}=\left[ ie^{\psi}a \sin\theta,0,e^{-\chi},\dfrac{i e^{-\psi}}{\sin\theta}\right]/\sqrt{2} (r+ia\cos \theta) .
\end{aligned}
\end{align}

We have not made any specification for the expressions of the functions $\psi$ and $\chi$, so this tetrad is exact for any metric of the form of Eq.~(\ref{eq:FKLkerr expo}). In the case of the KL metric, we know that these functions vanish when $q = 0$, so in this limit the tetrad reduces to the Kinnersley tetrad for the Kerr metric.

\subsection{Newman-Penrose Quantities for the Linear KL Metric}

Once the null tetrad is obtained, we proceed to compute the Newman-Penrose quantities for the KL metric. The explicit expressions of these quantities will be presented for the linear KL metric, which satisfies equation \ref{CHI}, and are valid up to order $O(q^{2}, qa, qM)$. However, this approximation is performed after having obtained the more general quantities, so the computational method presented is valid for the general KL metric with minor modifications when introducing approximations.

\subsection{Computation of the Quantities} \label{sec:aproxaprox}

We begin by taking the components of the KL metric from Eqs.~(\ref{KL}) in the linear limit (\ref{CHI}), the inverse metric, and the contravariant null tetrads Eqs.~(\ref{tetradaFKLnula}) are introduced, without specifying the values of the functions $\psi$ and $\chi$.

The definitions of the spin coefficients (Eqs.~(\ref{eq definicion coeficientes de spin})), Weyl (Eqs~\ref{eq:definicion weyl}) and Ricci scalars (Eqs.~(\ref{eq:definicion Ricci})) used in the calculations are presented in the appendix.

With all general quantities calculated, we performs the approximation. In the case of the linear KL metric, the condition $\psi = \chi$ is specified, and the derivatives of the function $\chi$ given by Eq.~(\ref{CHI}) so that

\begin{eqnarray}
    \dfrac{\partial\chi}{\partial r} &=& -\dfrac{3qP_{2}}{r^{4}} = qf_{1} ,\nonumber \\
    \dfrac{\partial\chi}{\partial \theta} &=& -\dfrac{3q\cos\theta\sin\theta}{r^{3}} = qf_{2}.
\end{eqnarray}

Then the approximations applied are

\begin{align}\label{eq:aproxxlinealh}
    q^{2}\sim 0\, \quad qa\sim0 \ , \quad qM\sim 0.
\end{align}

Therefore some expressions may be simplified

\begin{eqnarray}
    q\Delta &=& q(r^{2}-2Mr-a^{2})\approx qr^{2},  \nonumber \\
    q\Sigma &=& q(r^{2}+a^{2}\cos^{2}\theta)\approx qr^{2}.
\end{eqnarray}

\begin{align}
    q\varpi = q(r\Delta + 4rM^{2} + 2M\Delta)\approx qr\Delta\ ,
\end{align}

To obtain the expressions for the KL metric at quadratic order, the derivatives of $\psi$ and $\chi$ are specified based on their definitions of Eq.~(\ref{eq:FKL CUAD}), and the approximations

\begin{align}
    q^{3}, q^{2}a, qa^{2}, q^{2}M, qM^{2}, qaM \sim 0.
\end{align}

\subsubsection{Explicit Analytical Expressions}

At linear order, we obtain the following expressions for the spin coefficients:

\begin{eqnarray}
       \kappa &=& -\dfrac{6q\sin\theta\cos\theta}{\sqrt{2}\Delta\Sigma}e^{-\chi} , \nonumber \\ 
       \tau &=& \tau_{K}e^{-\chi}  , \nonumber \\ 
       \sigma &=& 0  , \nonumber \\ 
       \rho &=& \left[\rho_{K}+\dfrac{3q P_{2}}{r^{2}\Sigma}\right]e^{-\chi} , \nonumber \\ 
       \pi &=& \pi_{K}e^{-\chi}  , \nonumber \\ 
       \mu &=& \left[\mu_{K}+\dfrac{3q\Delta P_{2}}{2r^{2}\Sigma^{2}}\right]e^{-\chi} , \nonumber \\ 
       \nu &=& \dfrac{3q\Delta\sin\theta\cos\theta}{2\sqrt{2}\Sigma^{3}}e^{-\chi} ,  \\ 
       \lambda &=& 0 , \nonumber \\ 
       \varepsilon &=& \dfrac{3qP_{2}}{2r^{2}\Sigma}e^{-\chi} , \nonumber \\ 
       \alpha &=& \left[\alpha_{K}+\dfrac{3q\sin\theta\cos\theta}{2\sqrt{2}\Sigma^{2}}\right]e^{-\chi} \ ,\nonumber\\ 
       \beta &=& \left[\beta_{K}-\dfrac{3q\sin\theta\cos\theta}{2\sqrt{2}\Sigma^{2}}\right]e^{-\chi}  , \nonumber \\ 
       \gamma &=& \left[\gamma_{K}+\dfrac{3q\Delta P_{2}}{4r^{2}\Sigma^{2}}\right]e^{-\chi}  .\nonumber
\end{eqnarray}

\noindent and for the Weyl scalars we find

\begin{eqnarray}
    & \Psi_{0}=-\dfrac{3qr^{3}\sin^{2}\theta}{\Sigma^{3}\Delta}e^{-2\chi}  , \nonumber \\ 
    & \Psi_{1}=\dfrac{12qr^{3}\sin\theta\cos\theta}{\sqrt{2}\Sigma^{3}\Delta}e^{-2\chi}  , \nonumber \\ 
    & \Psi_{2}=\left[\Psi_{2K}-\dfrac{6qrP_{2}}{\Sigma^{3}}\right]e^{-2\chi}  ,  \\ 
    & \Psi_{3}=-\dfrac{6qr\sin\theta\cos\theta}{\sqrt{2}\Sigma^{3}}e^{-2\chi} , \nonumber \\ 
    & \Psi_{4}=-\dfrac{3qr\Delta\sin^{2}\theta}{4\Sigma^{4}}e^{-2\chi}. \nonumber
\end{eqnarray}

\noindent and for the Ricci scalars we obtain

\begin{eqnarray}
    \Phi_{00} &=& \Phi_{11} = \Phi_{22} = \Phi_{10} \nonumber \\ 
    &=& \Phi_{20} = \Phi_{12} = \Lambda = 0 .
\end{eqnarray}

We observe how all quantities reduce to the Kerr metric expressions in the limit $q=0$, which is also expected since the KL metric reduces to Kerr in this limit. In the limit $a=0$, the linear KL metric reduces to the linear order of the Erez--Rosen metric \cite{ERvsHT, frutos2016approximate}, redefining the parameter $q$, so we can consider the quantities in this limit as those describing the Erez--Rosen metric at linear order.

The terms proportional to the quadrupole mass moment parameter in the Weyl scalars have an asymptotic behavior, as $r \rightarrow \infty$, of

\begin{align}
    \Psi_{q} \sim \dfrac{1}{r^{5}}.
\end{align}

\noindent while $\Psi_{2K}$ has an asymptotic behavior of

\begin{align}
    \Psi_{2K} \sim \dfrac{M}{r^{3}} + \dfrac{3iMa}{r^{4}} .
\end{align}

We can reconstruct the Riemann tensor from the Weyl scalars and the Ricci scalars \cite{chandrasekhar1998mathematical}, which are zero in this case, so this asymptotic behavior indicates that as we move away from the compact object, spacetime can be divided into a region where the quadrupole mass moment parameter corrections are significant, and a region where the curvature behavior is practically the same as that of the Kerr solution.

\subsection{Petrov Type of the KL Metric}

Using the Weyl scalars of the KL metric, we can determine its Petrov type \cite{PedroTesis}. First, consider the special case where $\theta=0$ or $\theta=\pi$, which corresponds to the symmetry axis of the metric. We observe that in this case, all Weyl scalars vanish except for $\Psi_{2}$; moreover, the spin coefficients $\kappa$, $\sigma$, $\lambda$, and $\nu$ also vanish. Therefore, on the symmetry axis the solution satisfies the Goldberg-Sachs theorem \cite{wald2010general}, and we determine that the spacetime is type D in this region.

In the general case, no Weyl scalar vanishes, so we must study the multiplicities of the roots of equation Eq.~(\ref{eq:psi0petrov}) to determine the Petrov type of the metric \cite{letniowski1988improved}. We first observe that the Weyl scalars can be written as

\begin{eqnarray} \label{eq:psi0petrov}
    \Psi_0 z^4 + 4\Psi_1 z^3 + 6\Psi_2 z^2 + 4\Psi_3 z + \Psi_4 = 0 
\end{eqnarray}

\begin{eqnarray}
     \Psi_{0} &=& q \Psi_{0}^{(q)} e^{-2\chi}  , \nonumber \\ 
     \Psi_{1} &=& q \Psi_{1}^{(q)} e^{-2\chi}  , \nonumber \\ 
     \Psi_{2} &=& \left[ \Psi_{2K} + q \Psi_{2}^{(q)} \right] e^{-2\chi}  ,  \\ 
     \Psi_{3} &=& q \Psi_{3}^{(q)} e^{-2\chi}, \nonumber \\ 
     \Psi_{4} &=& q \Psi_{4}^{(q)} e^{-2\chi}. \nonumber 
\end{eqnarray}

Following the algorithm of Letniowski and McLenaghan \cite{letniowski1988improved}, for the case in which no scalar vanishes, we first calculate the quantity $H$ obtaining, to lowest order in $q$, the expression

\begin{align}
    H = \Psi_{4} \Psi_{2} - \Psi_{1}^{2} \approx q \Psi_{4}^{(q)} \Psi_{2K} e^{-4\chi} .
\end{align}

which indicates that this quantity does not vanish, so we proceed to calculate $I$ obtaining

\begin{eqnarray}
    I &=& 3 \Psi_{2}^{2} - 4 \Psi_{1} \Psi_{3} + \Psi_{0} \Psi_{4} \nonumber \\
    &\approx& 3 \Psi_{2K} (\Psi_{2K} + 2 q \Psi_{2}^{(q)}) e^{-4\chi} .
\end{eqnarray}

\noindent so this quantity also does not vanish. We then evaluate the quantity $G$ finding that it has the form

\begin{eqnarray}
    G &=& \Psi_{4}^{2} \Psi_{1} - 3 \Psi_{4} \Psi_{3} \Psi_{2} + 2 \Psi_{3}^{3} \nonumber \\
    &\approx& -3 q^{2} \Psi_{4}^{(q)} \Psi_{3}^{(q)} \Psi_{2K} e^{-6\chi}.
\end{eqnarray}

\noindent again this quantity is not zero, and we must then calculate $J$, obtaining

\begin{eqnarray}
    J &=& \Psi_{0} \Psi_{2} \Psi_{4} + 2 \Psi_{1} \Psi_{2} \Psi_{3} - \Psi_{0} \Psi_{3}^{2} - \Psi_{4} \Psi_{1}^{2} - \Psi_{2}^{3} \nonumber \\
    &\approx& - \Psi_{2K}^{2} (\Psi_{2K} + 3 q \Psi_{2}^{(q)}) e^{-6\chi}\ .
\end{eqnarray}

Finally, since $J$ is nonzero, we calculate the quantity $D$, which takes the following form

\begin{align}
    D = 27 J^{2} - I^{3} \approx -81 q^{2} \Psi_{0}^{(q)} \Psi_{4}^{(q)} \Psi_{2K}^{4} e^{-12 \chi} \ .
\end{align}

This last quantity is also nonzero, which indicates that the KL solution is of type I if terms of order $q^{2}$ are not neglected.

If we make the approximation $q^{2} \sim 0$, the quantity $G$ vanishes, so following the algorithm, we must calculate the quantity $Z$, obtaining

\begin{align}
    Z = I \Psi_{4}^{2} - 12 H^{2} \approx -9 q^{2} (\Psi_{4}^{(q)})^{2} \Psi_{2K}^{2} e^{-8 \chi} .
\end{align}

\noindent which also vanishes if quadratic terms in the quadrupolar parameter are neglected; therefore, in the linear regime, the KL metric can be considered approximately type D \cite{PedroTesis}. The previous results indicate that as the quadratic terms become relevant, the metric deviates from being approximately type D to being type I.

We can corroborate this statement by calculating the speciality index of the KL solution, obtaining

\begin{align}\label{eq:specialityKL}
    S = \dfrac{27 J^{2}}{I^{3}} \approx 1 - 3 q^{2} \dfrac{\Psi_{0}^{(q)} \Psi_{4}^{(q)}}{\Psi_{2K}^{2}}.
\end{align}

\noindent where it is indeed observed that the deviation of $S$ from unity is quadratic in $q$.

\begin{eqnarray*}\label{especialidadpert}
    S = 1 - 3 \epsilon^2 \frac{\Psi_1^{(1)} \Psi_4 ^{(1)} }{  \left(   \Psi_2 ^{(0)}    \right)^2  }  + O(\epsilon^3).
\end{eqnarray*}

The Eq.~(\ref{eq:specialityKL}) for the specialty index is equal to the index for a metric that is a perturbation of a type D solution (Eq.~\ref{especialidadpert}), in this case with $q$ as the perturbative parameter \cite{baker2000making}.

Explicitly, $S$ takes the form:

\begin{align}
    S \approx 1 - q^{2} \dfrac{27 r^{4} \sin^{4} \theta (r - i a \cos \theta)^{6}}{4 M^{2} \Sigma^{7}} .
\end{align}

\noindent which, for large values of $r$ has the following asymptotic behavior

\begin{align}
    S \approx 1 - q^{2} \dfrac{27 \sin^{4} \theta}{4 M^{2} r^{4}} .
\end{align}

This expression is the same as that obtained for the Hartle--Thorne metric in this limit \cite{berti2005rotating}, if we make the substitution $q \rightarrow J^{2}/M - Q$, which corresponds to the substitution made in \cite{frutos2016approximate} to relate the Hartle--Thorne solution with the KL solution. We also observe that $S$ is exactly equal to $1$ on the symmetry axis, which is expected since in this region the metric is exactly type D. This behavior is the same as that of the Erez--Rosen solution, which is type I except on the axis, where it is type D \cite{PedroTesis}.

Finally, we emphasize that the previous results were obtained using the KL metric at linear order; however, if we use the more general expressions of the KL solution, the only difference would be that we would obtain terms proportional to $q^{2}$, $q a$, and $q M$ in the Weyl scalars, which would help specify the next orders of $q$ in the deviation of $S$ from unity. However, the Petrov behavior would remain the same — type I in general and approximately type D at linear order. For this reason, explicit expressions at this order are not presented, as they would not contribute relevant results for the research \cite{PedroTesis}.

\section{ \label{ch:PFKL}    Perturbations of the Linear KL Solution  }

Having the NP description of the KL metric, we can now perform gravitational, electromagnetic, and scalar perturbations within this formalism. Moreover, we know that the linear-order KL metric is approximately of type D, so we can carry out perturbations based on the Teukolsky formalism for exact type D metrics, making the necessary approximations \cite{PedroTesis}.

\subsection{Special Relations for the KL Metric}\label{sec:especiales}

In order to use the Teukolsky perturbation formalism, we must verify that the linear-order KL metric satisfies the relations for type D metrics. Since the KL metric is a vacuum solution, we know that the following holds

\begin{align}
    \Phi_{00} =\Phi_{11} =\Phi_{22} =\Phi_{10} =\Phi_{20} =\Phi_{21} =\Lambda =0.
\end{align}

\noindent so the conditions for the Ricci scalars pose no issue.

It also holds that

\begin{align}
    \sigma=\lambda=0.
\end{align}

\noindent but we observe that the Weyl scalars $\Psi_{0}$, $\Psi_{1}$, $\Psi_{3}$, and $\Psi_{4}$, as well as the spin coefficients $\kappa$ and $\nu$, do not vanish, even at linear order in $q$. These will contribute terms to the perturbation equations of the form

\begin{align}
    A^{(0)}B^{(1)}
\end{align}

\noindent where $A$ is one of the aforementioned quantities, and $B$ is any NP quantity. However, we note that the Weyl scalars and spin coefficients that should vanish are proportional to $q$, which is considered a small first-order parameter in the linear approximation of the KL metric, and the quantities $B^{(1)}$ are first-order in the perturbation. Therefore, these terms are proportional to

\begin{align}
   qB^{(1)}
\end{align}

\noindent which is a second-order quantity and can thus be neglected in the first-order perturbation equations.

Therefore, our first approximation is

\begin{align}\label{eq:aproxfkl1}
    qB^{(1)}\sim 0
\end{align}

\noindent which indicates that the perturbation equations will be valid as long as the perturbative fields are of the same order as the corrections to the gravitational field of a compact object given by the quadrupole parameter $q$ at linear order.

For a vacuum type D metric satisfies \cite{teukolsky1973perturbations}

\begin{eqnarray}
    \Psi_0 &=& \Psi_1 = \Psi_3 = \Psi_4 = 0, \nonumber \\
    \kappa &=& \sigma = \lambda = \nu = 0.
\end{eqnarray}

Therefore, the Bianchi identities of Eqs.~(\ref{eq:bianchi5}), (\ref{eq:bianchi6}), (\ref{eq:bianchi7}) and (\ref{eq:bianchi8}) simplifies to \cite{teukolsky1973perturbations}

\begin{eqnarray}\label{eq:bianchivacuum}
    D\Psi_2 &=& 3 \rho \Psi_2, \nonumber \\
    \Delta \Psi_2 &=& -3 \mu \Psi_2, \nonumber \\
    \bar{\delta} \Psi_2 &=& -3 \pi \Psi_2, \nonumber \\
    \delta \Psi_2 &=& 3 \tau \Psi_2. 
\end{eqnarray}

On the other hand, for the linear KL metric, the same Bianchi identities reduce, after neglecting terms of order $q^{2}$, to

\begin{eqnarray}\label{eq:bianchifkl}
    D\Psi_{2} &=& 3\rho\Psi_{2}+2(\pi-\alpha)\Psi_{1} + \bar{\delta}\Psi_{1} , \nonumber  \\
    \Delta\Psi_{2} &=& -3\mu\Psi_{2}+2(\beta-\tau)\Psi_{3}+\delta\Psi_{3}  ,   \\
    \bar{\delta}\Psi_{2} &=& -3\pi\Psi_{2}+2(\varepsilon-\rho)\Psi_{3}+D\Psi_{3}  , \nonumber  \\
    \delta\Psi_{2} &=& 3\tau\Psi_{2}+2(\mu-\gamma)\Psi_{1}+\Delta\Psi_{1}.  \nonumber  
\end{eqnarray}

The extra terms in these relations, compared to Eqs.~(\ref{eq:bianchivacuum}) for exact type D metrics, are again proportional to $q$, and also contribute with terms proportional to $qB^{(1)}$ in the perturbation equations. Therefore, the approximation (\ref{eq:aproxfkl1}) is sufficient to eliminate these contributions.

To obtain decoupled perturbation equations in the Teukolsky formalism, it is necessary to use the commutation relation \cite{teukolsky1973perturbations}

\begin{eqnarray}\label{CONMUTACIÓNOSOM}
    [D - (u+1)\varepsilon+\bar{\varepsilon} +v\rho-\bar{\rho}](\delta-u\beta+v\tau)   \nonumber \\
    -[\delta - (u+1)\beta-\bar{\alpha}+\bar{\pi}+v\tau](D-u\varepsilon+v\rho)=0 .
\end{eqnarray}

\noindent which holds for exact type D solutions. In the case of the KL metric, evaluating this commutation relation explicitly for an arbitrary quantity $\Theta$ and we find that the relation does not hold. Instead, the result is proportional to terms of the form

\begin{align}\label{eq:extraderivadas}
    q\Theta,\ q\dfrac{\partial \Theta}{\partial t}\ , \ q\dfrac{\partial \Theta}{\partial r} .
\end{align}

Nevertheless, we recall that this relation is applied to first-order quantities in the perturbation ($\Theta = B^{(1)}$), so if we make the approximations

\begin{eqnarray}\label{eq:aproxderivadas}
        q\dfrac{\partial B^{(1)}}{\partial t}\sim0 ,\\
        q\dfrac{\partial B^{(1)}}{\partial r}\sim0 .   \nonumber 
\end{eqnarray}

\noindent along with the approximation (\ref{eq:aproxfkl1}), the relation (\ref{CONMUTACIÓNOSOM}) is approximately satisfied, and we can neglect the extra terms of the form (\ref{eq:extraderivadas}) in the perturbation equations. Physically, this indicates that the perturbation equations will be valid for perturbative fields that vary slowly both in time and with distance as one moves away from the compact object \cite{PedroTesis}.

Finally, we observe that the proposed approximations, being applied equally to all first-order quantities, do not break the symmetry of the NP equations under the transformation $l^{\mu} \longleftrightarrow n^{\mu}$, $m^{\mu} \longleftrightarrow \bar{m}^{\mu}$, so we can continue employing this transformation to derive new equations in the same manner used by Teukolsky \cite{PedroTesis}.

\subsection{Perturbation Equations}

The equations for the first-order perturbative fields of the linear KL metric are derived using the same procedure employed by Teukolsky for vacuum perturbations of type D metrics, so any extra terms in the equations—induced by the fact that the type D conditions are not exactly satisfied—will be neglected under the approximations introduced in the previous section \cite{PedroTesis}.

This means that the perturbative equations for the linear KL metric will take the same form as those for an exact type D metric, with the caveat that they will be approximate in the case of the KL solution, except for the scalar field equation, which is exact for general spacetimes. Therefore, the equations for the fields $\Psi_{0}^{(1)}$, $\Psi_{4}^{(1)}$, $\phi_{0}^{(1)}$, $\phi_{2}^{(1)}$, and $\Phi^{(1)}$ will be given by equations \cite{teukolsky1973perturbations}

\begin{itemize}
    \item For gravitational fields

    \begin{eqnarray}
        [(D &-& 4\rho - \bar{\rho}-3\varepsilon+\bar{\varepsilon})(\Delta-4\gamma+\mu) \nonumber \\
        &-&(\delta-4\tau+\bar{\pi}-\bar{\alpha}-3\beta)(\bar{\delta}-4\alpha+\pi) \nonumber \\
        &-& 3\Psi_{2} ] \Psi_{0}^{(1)}=4\pi T_{0}^{(1)}\ ,
    \end{eqnarray}

    \begin{eqnarray}
        [(\Delta &+& 4\mu+\bar{\mu}+3\gamma-\bar{\gamma})(D+4\varepsilon-\rho)  \nonumber\\
        &-& (\bar{\delta}+4\pi-\bar{\tau}+\bar{\beta}+3\alpha)(\delta+4\beta-\tau)  \nonumber \\
        &-& 3\Psi_{2}]\Psi_{4}^{(1)}=4\pi T_{4}^{(1)} .
    \end{eqnarray}

    \item For electromagnetic fields

    \begin{eqnarray}
        [(D &-& \varepsilon + \bar{\varepsilon} - 2\rho - \bar{\rho})(\Delta+\mu-2\gamma) \nonumber\\
         &-& (\delta-\beta-\bar{\alpha}-2\tau+\bar{\pi})(\bar{\delta}+\pi \nonumber \\
         &-& 2\alpha)]\phi_{0}^{(1)}=2\pi J_{0}^{(1)} ,
    \end{eqnarray}

    \begin{eqnarray}
        [(\Delta &+& \gamma-\bar{\gamma}+2\mu +  \bar{\mu})(D-\rho+2\varepsilon)  \nonumber \\
         &-& (\bar{\delta}+\alpha+\bar{\beta}+2\pi-\bar{\tau})(\delta  -\tau \nonumber  \\
         &+& 2\beta)]\phi_{2}^{(1)}=2\pi J_{2}^{(1)} .
    \end{eqnarray}

    \item For scalar fields

    \begin{eqnarray}
        [(D &+& \varepsilon  +\bar{\varepsilon}-\bar{\rho}-\rho)\Delta  \nonumber \\
        &-& (\delta+\beta-\bar{\alpha}+\bar{\pi}-\tau)\bar{\delta}  \nonumber   \\
        &+& (\Delta-\gamma-\bar{\gamma}+\mu+\bar{\mu})D  \nonumber \\
        &-& (\bar{\delta}-\alpha+\bar{\beta}-\bar{\tau}+\pi)\delta]\Phi^{(1)}=4\pi T^{(1)}\ .
    \end{eqnarray}

\end{itemize}

In these equations, the index $(0)$ for the background quantities is understood, and the perturbation sources $T_{0}^{(1)}$, $T_{4}^{(1)}$, $J_{0}^{(1)}$, $J_{2}^{(1)}$ are given by

\begin{eqnarray}
    T_{0}^{(1)}  &=& (\delta-4\tau+\bar{\pi}-\bar{\alpha}-3\beta) \bigg[(D-2\varepsilon-2\bar{\rho})T_{lm}^{(1)}   \nonumber \\
    &-& (\delta+\bar{\pi}-2\bar{\alpha}-2\beta)T_{ll}^{(1)} \bigg] \nonumber  \\
    &+&(D-4\rho-\bar{\rho}-3\varepsilon+\bar{\varepsilon})\bigg[(\delta+2\bar{\pi}-2\beta)T_{lm}^{(1)}  \nonumber \\
    &-& (D-2\varepsilon+2\bar{\varepsilon}-\bar{\rho})T_{mm}^{(1)}\bigg]\ ,
\end{eqnarray}

\begin{eqnarray}
    T_{4}^{(1)} &=& (\bar{\delta} + 4\pi-\bar{\tau}+\bar{\beta}+3\alpha) \bigg[(\Delta+2\gamma+2\bar{\mu})T_{n\bar{m}}^{(1)} \nonumber \\
    &-& (\bar{\delta}-\bar{\tau}+2\bar{\beta}+2\alpha)T_{nn}^{(1)} \bigg] \nonumber \\
    &+& (\Delta+4\mu+\bar{\mu}+3\gamma-\bar{\gamma})\bigg[(\bar{\delta}-2\bar{\tau}+2\alpha)T_{n\bar{m}}^{(1)} \nonumber \\
    &-& (\Delta+2\gamma-2\bar{\gamma}+\bar{\mu})T_{\bar{m}\bar{m}}^{(1)}\bigg ]\ ,
\end{eqnarray}

\begin{eqnarray}
    J_{0}^{(1)} &=& (\delta-\beta-\bar{\alpha}-2\tau+\bar{\pi})J_{l}^{(1)}  \nonumber \\
                &-& (D-\varepsilon+\bar{\varepsilon}-2\rho-\bar{\rho})J_{m}^{(1)}\ ,
\end{eqnarray}

\begin{eqnarray}
     J_{2}^{(1)} &=& (\Delta+\gamma-\bar{\gamma}+2\mu+ \bar{\mu})J_{\bar{m}}^{(1)}  \nonumber \\
                 &-&  (\bar{\delta}+\alpha+\bar{\beta}+2\pi-\bar{\tau})J_{n}^{(1)}\ .
\end{eqnarray}

\subsection{Master Equation for the KL Metric}

The perturbation equations are evaluated explicitly in a coordinate system with the help of a REDUCE computer algebra system program, applying the NP quantities for the KL metric found previously and implementing the approximations described above \cite{PedroTesis}. The program also follows the methodology presented in \ref{sec:aproxaprox} to introduce the linear approximations of the KL metric and ends by simplifying the numerators of the differential equations for each field.

We find that the perturbation equations can be written as a single \textit{Master} equation (TME) for the KL metric:

\begin{eqnarray}\label{eq:MASTERFKL}
    &e^{4\chi}&\left[\dfrac{(r^{2}+a^{2})^{2}}{\Delta}-a^{2}\sin^{2}\theta\right]\dfrac{\partial^{2}\psi}{\partial t^{2}} \nonumber \\
    &+& e^{2\chi}\dfrac{4Mar}{\Delta}\dfrac{\partial^{2}\psi}{\partial t\partial\varphi}+\left[\dfrac{a^{2}}{\Delta}-\dfrac{1}{\sin^{2}\theta}\right]\dfrac{\partial^{2}\psi}{\partial \varphi^{2}} \nonumber \\  
    &-&\Delta^{-s}\dfrac{\partial}{\partial r}\left(\Delta^{s+1}\dfrac{\partial\psi}{\partial r}\right)-\dfrac{1}{\sin\theta}\dfrac{\partial}{\partial\theta}\left(\sin\theta\dfrac{\partial\psi}{\partial \theta}\right) \nonumber \\
    &-& 2s\left[\dfrac{a(r-M)}{\Delta}+\dfrac{i\cos\theta}{\sin^{2}\theta}\right]\dfrac{\partial\psi}{\partial\varphi}  \nonumber \\  
    &-& 2se^{2\chi}\left[\dfrac{M(r^{2}-a^{2})}{\Delta}-r-ia\cos\theta \right]\dfrac{\partial\psi}{\partial t} \nonumber \\
    &+& (s^{2}\cot^{2}\theta-s)\psi + s\dfrac{6qi\cos\theta}{r^{3}}\dfrac{\partial\psi}{\partial\varphi} \nonumber \\
    &=& 4e^{2\chi}\pi\Sigma T .
\end{eqnarray}

\noindent where $M$, $a$, $\Delta$, $\Sigma$, and $\chi$ have their usual meanings for the linear KL metric, and $s$ is the \textit{spin} parameter. Together with the field $\psi$ and its respective source $T$, they can take the values indicated in Table \ref{tabla:masterfkl}.

\begin{table}[t]
    \centering
    \renewcommand{\arraystretch}{1.45}
    \setlength{\tabcolsep}{14pt} 
    \large 
    \begin{tabular}{c c c}
        \hline
        $\boldsymbol{\psi}$ & $\boldsymbol{s}$ & $\boldsymbol{T}$ \\
        \hline
        $\Phi^{(1)}$
        & $0$
        & $T^{(1)}$ \\
        \hline
        $\phi_{0}^{(1)}$
        & $1$
        & $J_{0}^{(1)}$ \\[2pt]
        
        $\rho_{K}^{-2}\phi_{2}^{(1)}$
        & $-1$
        & $\rho_{K}^{-2}J_{2}^{(1)}$ \\
        \hline
        $\Psi_{0}^{(1)}$
        & $2$
        & $2T_{0}^{(1)}$ \\[2pt]
        
        $\rho_{K}^{-4}\Psi_{4}^{(1)}$
        & $-2$
        & $2\rho_{K}^{-4}T_{4}^{(1)}$ \\
        \hline
    \end{tabular}
    \caption{$\psi$, $s$, and $T$ for the TME of the KL metric.}
    \label{tabla:masterfkl}
\end{table}

We observe that the equation for the perturbations differs from the TME (Eq.~(\ref{EQ:MASTERKERR})) in the exponential factors that depend on $\chi$, and contains an extra term depending on $q$, so that in the limit $q = 0$ the equation correctly reduces to the Kerr case. For $a = 0$, the equation describes perturbations around a static deformed object; that is, perturbations of the linear-$q$ limit of the Erez--Rosen solution.

\subsubsection{Separability of the Equation}

Despite the similarity between the \textit{Master} equations for the KL and Kerr metrics, the presence of the exponential factors and the extra term have relevant consequences for the separability of the homogeneous equation ($T = 0$). These contributions, which depend on $q$, are functions $f$ of $r$ and $\theta$, such that

\begin{align}
    f(r,\theta)\neq f_{1}(r)+f_{2}(\theta) ,
\end{align}

\noindent so it will not be possible to separate the terms into functions that depend solely on $r$ and solely on $\theta$, and we will not be able to obtain separate radial and angular equations.

Nevertheless, we can obtain a coupled radial-angular equation that will not depend on $t$ and $\varphi$. For this, we write the field $\psi$ as

\begin{align}
    \psi = e^{-i\omega t}e^{im\varphi}F(r,\theta) .
\end{align}

\noindent and find that Eq.~(\ref{eq:MASTERFKL}) reduces to the following equation for $F$

\begin{eqnarray}\label{eq:angularradial}
    & & \bigg\{ - \Delta^{-s}\dfrac{\partial}{\partial r}\left(\Delta^{s+1}\dfrac{\partial }{\partial r}\right)-\dfrac{1}{\sin\theta}\dfrac{\partial}{\partial\theta}\left(\sin\theta\dfrac{\partial }{\partial \theta}\right)  \nonumber\\
    &+& m\omega e^{2\chi}\dfrac{4Mar}{\Delta} - \omega^{2} e^{4\chi}\left[\dfrac{(r^{2}+a^{2})^{2}}{\Delta}-a^{2}\sin^{2}\theta\right] \nonumber \\    
   &-& 2ims\left[\dfrac{a(r-M)}{\Delta}+\dfrac{i\cos\theta}{\sin^{2}\theta}-\dfrac{3qi\cos\theta}{r^{3}}\right]  \nonumber  \\
   &-& m^{2}\left[\dfrac{a^{2}}{\Delta}-\dfrac{1}{\sin^{2}\theta}\right]  \nonumber \\
   &+& 2i\omega se^{2\chi}\left[\dfrac{M(r^{2}-a^{2})}{\Delta}-r-ia\cos\theta \right]  \nonumber \\
   &+& s^{2}\cot^{2}\theta-s\bigg\}F = 0 .
\end{eqnarray}

The above equation, together with the regularity conditions at $\theta=0$ and $\theta=\pi$, and the appropriate boundary conditions for the behavior in $r$ for a given specific problem, will determine the behavior of the perturbations in the exterior of a rotating deformed compact object.

We thus observe how the quadrupole parameter introduces a significant difference in the behavior of the perturbation equation compared to that of the Kerr metric, since the presence of $q$ induces coupling between the angular and radial parts of the solutions to the homogeneous equation \cite{PedroTesis}.

\subsubsection{Asymptotic Behavior}

As in the case of the TME for Kerr, we can analyze the asymptotic behavior of the field $\psi$ without solving the perturbation equation exactly. To do so, we note that

\begin{align}\label{eq:exponencialestaylor}
    e^{b\chi}=e^{bqP_{2}/r^{3}}= 1+\dfrac{bq}{r^{3}}P_{2} + O(r^{-6})\ ,
\end{align}

where $b$ is an arbitrary constant. Therefore, the exponential factors, as well as the extra $q$-dependent term in Eq.~(\ref{eq:MASTERFKL}), decay as $r^{-3}$ in the limit $r\rightarrow\infty$. On the other hand, the terms independent of $q$, that is, those that match the Kerr equation, decay as $r^{-2}$ or more slowly in this limit.

The above analysis indicates that asymptotically, the terms depending on the quadrupole parameter can be neglected, and the perturbation equation for the KL metric will tend to that of the Kerr metric as $r\rightarrow\infty$. Therefore, the asymptotic behavior of the various perturbative fields will be obtain following a similar procedure used by \cite{teukolsky1973perturbations} 

\begin{itemize}
    \item for outgoing waves

    \begin{eqnarray*}
        \Phi^{(1)},\phi_{2}^{(1)},\Psi_{4}^{(1)}\sim \dfrac{e^{i\omega r*}}{r}, \\
        \phi_{0}^{(1)}\sim \dfrac{e^{i\omega r*}}{r^{3}},\quad \Psi_{0}^{(1)}\sim \dfrac{e^{i\omega r*}}{r^{5}}\ ;
    \end{eqnarray*}

    \item and for ingoing waves

    \begin{eqnarray*}
         \Phi^{(1)},\phi_{0}^{(1)},\Psi_{0}^{(1)}\sim \dfrac{e^{-i\omega r*}}{r}, \\
         \phi_{2}^{(1)}\sim \dfrac{e^{-i\omega r*}}{r^{3}},\quad \Psi_{4}^{(1)}\sim \dfrac{e^{-i\omega r*}}{r^{5}}\ .
    \end{eqnarray*}

\end{itemize}

In the case of the limit $r\rightarrow r_{+}$, we observe that several terms in Eq.~(\ref{eq:MASTERFKL}) that are inversely proportional to $\Delta$, and thus diverge and correspond to the dominant contributions in the differential equation, include exponential factors. Therefore, we cannot reduce the KL equation to the Kerr one, nor separate it in this limit.

However, we must recall that the equation is valid for the exterior of a deformed compact object, and due to the no-hair conjecture, a black hole cannot possess a quadrupole moment. Thus, objects described by the KL solution will not exhibit an event horizon, and their surface will be located at $r>r_{+}$, meaning that the limit $r\rightarrow r_{+}$ will be irrelevant for the TME of the KL metric. Instead, boundary conditions must be specified at the object's surface \cite{PedroTesis}.

\subsubsection{Solution to the Coupled Radial-Angular Equation}

Although Eq.~(\ref{eq:angularradial}), which is coupled in the radial and angular parts, is not exactly separable, we can use approximations to solve it. First, recall that in deriving the TME for KL metric, we used the approximation

\begin{align*}
    q\dfrac{\partial \psi}{\partial t}\sim0 ,
\end{align*}

and that upon separating the equation in the $t$ variable, the time dependence of the field $\psi$ has the form

\begin{align*}
    \psi\sim e^{-i\omega t} .
\end{align*}

Therefore,

\begin{align}
    \dfrac{\partial \psi}{\partial t}\sim \omega e^{-i\omega t} ,
\end{align}

so the approximation on the time derivative implies that the radial-angular equation is only valid when

\begin{align}\label{eq:frecuenciaapprox}
    q\omega\sim 0\ . 
\end{align}

On the other hand, we note that the exponential factors in Eq.~(\ref{eq:angularradial}) appear with $\omega$ in the form

\begin{align}
    \omega e^{2\chi} ,
\end{align}

so, expanding the exponential (Eq.~(\ref{eq:exponencialestaylor})) and using approximation (\ref{eq:frecuenciaapprox}), we obtain

\begin{align}
    \omega e^{2\chi}\approx\omega\left(1+q\dfrac{2P_{2}}{r^{3}}\right)\approx \omega .
\end{align}

In this way, the exponential factors disappear from Eq.~(\ref{eq:angularradial}), which becomes

\begin{eqnarray}\label{qe:radialangularsinexponenciales}
    \bigg\{ &-& \Delta^{-s}\dfrac{\partial}{\partial r}\left(\Delta^{s+1}\dfrac{\partial }{\partial r}\right)-\dfrac{1}{\sin\theta}\dfrac{\partial}{\partial\theta}\left(\sin\theta\dfrac{\partial }{\partial \theta}\right) \nonumber \\
    &+& m\omega \dfrac{4Mar}{\Delta} - \omega^{2} \left[\dfrac{(r^{2}+a^{2})^{2}}{\Delta}-a^{2}\sin^{2}\theta\right]  \nonumber \\
    &-& 2ims\left[\dfrac{a(r-M)}{\Delta}+\dfrac{i\cos\theta}{\sin^{2}\theta}-\dfrac{3qi\cos\theta}{r^{3}}\right]  \nonumber \\
    &-&  m^{2}\left[\dfrac{a^{2}}{\Delta}-\dfrac{1}{\sin^{2}\theta}\right]   \nonumber\\
    &+& i\omega 2s\left[\dfrac{M(r^{2}-a^{2})}{\Delta}-r-ia\cos\theta \right] \nonumber \\
    &+& s^{2}\cot^{2}\theta-s\bigg\}F=0 .
\end{eqnarray}

The term depending on the quadrupole mass moment parameter is then the only non-separable term in the equation. Furthermore, since the exponential factors are absent, the remaining terms correspond exactly to those of the radial-angular equation for Kerr perturbations before separation into the radial and angular equations \cite{teukolsky1973perturbations}.

We can then write Eq.~(\ref{qe:radialangularsinexponenciales}) as

\begin{align}
    (\mathcal{L}_{K}+q\mathcal{L}_{q})F=0 ,
\end{align}

\noindent where $\mathcal{L}_{K}$ corresponds to the radial-angular operator for perturbations in the Kerr metric, and $\mathcal{L}_{q}$ is the additional term

\begin{align}
    \mathcal{L}_{q}=-\dfrac{6ms\cos\theta}{r^{3}} .
\end{align}

Proposing that the function $F$ has the form

\begin{align}
    F=F_{K} + qf\ ,
\end{align}

\noindent where $F_{K}$ is a solution to the Kerr metric equation,

\begin{align}\label{eq:operadorescero}
    \mathcal{L}_{K}F_{K}=0 ,
\end{align}

\noindent and expanding the equation, we obtain

\begin{align}
    \mathcal{L}_{K}F_{K}+q\mathcal{L}_{q}F_{K}+q\mathcal{L}_{K}f+q^{2}\mathcal{L}_{q}f=0 .
\end{align}

Neglecting the quadratic term in $q$, and using condition of Eq.~(\ref{eq:operadorescero}), we finally obtain an equation for the function $f$

\begin{align}
  \mathcal{L}_{K}f=-\mathcal{L}_{q}F_{K}.
\end{align}

The above equation corresponds to the radial-angular perturbation equation of the Kerr metric, which is separable and whose solutions form a complete set, with a source term. This means that we can construct $f$ as a particular solution using the eigenfunctions of the homogeneous equation for the Kerr metric \cite{PedroTesis}.

\subsection{Limitations of the Equation}

It is important to emphasize that the TME for perturbations of the KL metric was derived using several approximations, both in the metric and in the perturbative fields, such that its applicability is limited to the conditions imposed by these approximations.

First, the derivation is valid only for the linear KL metric, since only in this regime can the KL solution be approximately considered as type D. If quadratic terms in the quadrupole parameter are taken into account, the metric becomes type I, and it is no longer possible to find decoupled equations for the perturbative fields. This means that for compact objects modeled as highly oblate spheroids, Eq.~(\ref{eq:MASTERFKL}) will not provide a correct description of the perturbations \cite{PedroTesis}.

We also recall that the approximations indicated in subsection \ref{sec:especiales} imply that Eq.~(\ref{eq:MASTERFKL}) will not provide an adequate description for perturbations whose order of magnitude is not similar to the order of the gravitational field corrections due to the presence of $q$, nor for perturbations that vary rapidly with respect to the coordinates $t$ and $r$.

Likewise, the approximation in the time derivative implies that high-frequency modes $\omega$ are not correctly described by Eq.~(\ref{eq:angularradial}), and that the resolution method for the coupled radial-angular equation presented in the previous section is only valid for low-frequency modes \cite{PedroTesis}.

We emphasize that in the derivation of the scalar perturbation equation, no approximations were needed other than considering the scalar field as a test field. Therefore, in this case, Eq.~(\ref{eq:MASTERFKL}) is exact for the linear KL metric, and the coupled radial-angular equation adequately describes any mode with arbitrary $\omega$. However, the resolution method presented in the previous section is still only valid for low-frequency modes \cite{PedroTesis}.

Finally, we recall that the equation describes perturbations in the exterior of a compact object and will be subject to boundary conditions that must be specified at the surface of the object. However, in order to specify these conditions for a realistic astrophysical problem, it is necessary to know the description of the perturbations in the interior of the object, which is a more complex problem and beyond the scope of this work, so that there is an appropriate matching of the perturbation values at the boundary between both regions \cite{PedroTesis}.

\section{Conclusions} \label{Sec:Conclusions}

In principle, since this is an extension of TME, the KL equation can be used to include the influence of the quadrupole parameter in applications involving perturbations of the Kerr metric, thereby extending the results to provide a more realistic description for models of compact objects.

Among the applications of TME for the KL metric are: the study of the stability of the Kerr metric \cite{press1973perturbations}, the scattering of gravitational and electromagnetic waves, as well as superradiance phenomena \cite{teukolsky1974perturbations}, the production of gravitational and electromagnetic radiation by particles falling into a compact object \cite{degollado2014electromagnetic}, and accretion processes \cite{chitre1975electromagnetic}, the study of the electromagnetic spectrum of relativistic \textit{jets} \cite{staicova2011spectrum}, and the determination of the quasi-normal modes of compact objects \cite{kokkotas1999quasi}, among others.

It is worth noting that the extension of the above applications is restricted by the validity of the approximations used in the derivation of Eq.~(\ref{eq:MASTERFKL}), as well as by the proper specification of boundary conditions at the surface of the compact objects under study.

Moreover, the aforementioned applications are merely a guide to some of the possible phenomena that can be analyzed through perturbations. However, since compact objects remain an active area of research, new applications may emerge in the future.


\newpage

\begin{widetext}

\section{Appendix}


\section*{\label{ch:NPform} Newman-Penrose Formalism of General Relativity  }

Commonly, general relativity is treated using a coordinate basis to represent tensor components. However, there are different ways to represent the theory’s equations, one of them being the one developed by Ezra Newman and Roger Penrose in 1962 \cite{newman1962approach}, known as the \textit{Newman-Penrose formalism} (NP). This formalism will be presented as a special case of a more general formalism known as the \textit{tetrad formalism}, in which a non-coordinate basis, i.e., independent of the coordinates, is used to represent tensor components.

\subsection{Tetrad formalism}

In the tetrad formalism, at each point of spacetime, a set of four linearly independent vectors $\hat{e}_{(a)}^{\ \mu}$ is chosen, known as a \textit{tetrad}, which will serve as a basis \cite{chandrasekhar1998mathematical}. The indices in parentheses are the \textit{tetrad indices}, taking values $(0,1,2,3)$ and identifying each of the four basis vectors, while the Greek indices indicate components in the coordinate basis. These vectors will have associated co-vectors $\hat{e}_{(a)\mu}=g_{\mu\nu}\hat{e}_{(a)}^{\ \nu}$.

There is also a definition of an \textit{inverse tetrad} $\hat{e}^{(b)}_{\ \mu}$ such that it satisfies \cite{chandrasekhar1998mathematical}

\begin{eqnarray}
    \hat{e}_{(a)}^{\ \mu}\hat{e}^{(b)}_{\ \mu} &=& \delta_{(a)}^{(b)} , \nonumber \\
    \quad \hat{e}_{(a)}^{\ \mu}\hat{e}^{(a)}_{\ \nu} &=& \delta_{\nu}^{\mu}.
\end{eqnarray}

Likewise, the tetrad satisfies the condition

\begin{eqnarray}
   \hat{e}_{(a)}\!^{\mu}\hat{e}_{(b)\mu} = g_{\mu\nu}\hat{e}_{(a)}\!^{\mu}\hat{e}_{(b)}\!^{\nu} = \eta_{(a)(b)}.
\end{eqnarray}

\noindent where $\eta_{(a)(b)}$ is a constant symmetric matrix, which has an inverse matrix $\eta^{(a)(b)}$ such that

\begin{eqnarray}
    \eta^{(a)(b)}\eta_{(b)(c)} = \delta^{(a)}_{(c)}.
\end{eqnarray}

From these conditions it follows that \cite{chandrasekhar1998mathematical}

\begin{eqnarray}
    \eta_{(a)(b)}\hat{e}^{(a)}_{\ \mu} &=& \hat{e}_{(b)\mu}, \nonumber \\
    \eta^{(a)(b)}\hat{e}_{(a)\mu} &=& \hat{e}^{(b)}_{\ \mu},\\
    \hat{e}_{(a)\mu}\hat{e}^{(a)}_{\ \nu} &=& \eta_{(a)(b)}\hat{e}^{(a)}_{\ \mu}\hat{e}^{(a)}_{\ \nu} = g_{\mu\nu}.  \nonumber 
\end{eqnarray}

\noindent and we can project tensors and vectors to obtain their components with respect to the tetrad, such that

\begin{eqnarray}
    V^{(a)} &=& \hat{e}^{(a)}_{\ \mu}V^{\mu}, \nonumber   \\
    V^{\mu} &=& \hat{e}_{(a)}^{\ \mu}V^{(a)},    \\
    T_{(a)(b)} &=& \hat{e}_{(a)}\!^{\mu}\hat{e}_{(b)}\!^{\nu}T_{\mu\nu},  \nonumber  \\
    T_{\mu\nu} &=& \hat{e}^{(a)}_{\ \mu}\hat{e}^{(b)}_{\ \nu}T_{(a)(b)}.\label{eq:tcomp}   \nonumber 
\end{eqnarray}

From the previous equations we observe that the matrix $\eta_{(a)(b)}$ is simply the projection of the metric $g_{\mu\nu}$ onto the tetrad, and we can use this matrix to raise and lower tetrad indices \cite{chandrasekhar1998mathematical}. Likewise, since the tetrad is defined independently of coordinates, the tensor components in this basis behave as scalars with respect to coordinate transformations and derivatives \cite{alcubierre2008introduction}.

Following \cite{alcubierre2008introduction}, we now define directional derivatives along the tetrad as

\begin{eqnarray}
    A_{(a),(b)} &=& \hat{e}_{(b)}\!^{\mu}\partial_{\mu}A_{(a)} = \hat{e}_{(b)}\!^{\mu}\partial_{\mu}(\hat{e}_{(a)}\!^{\nu}A_{\nu}) \nonumber  \\
    &=& \hat{e}_{(b)}\!^{\mu}\nabla_{\mu}(\hat{e}_{(a)}\!^{\nu}A_{\nu}),
\end{eqnarray} 

\noindent so that

\begin{eqnarray}
    A_{(a),(b)} &=& \hat{e}_{(a)}\!^{\nu}\hat{e}_{(b)}\!^{\mu}\nabla_{\mu}A_{\nu}  \nonumber \\  
                &+& \hat{e}_{(b)}\!^{\mu}\hat{e}_{(c)}\!^{\nu}A^{(c)}\nabla_{\mu}\hat{e}_{(a)\nu}.
\end{eqnarray}

We also define the \textit{Ricci rotation coefficients}

\begin{align}
    \gamma_{(a)(b)(c)}=\hat{e}_{(a)}\!^{\mu}\hat{e}_{(c)}\!^{\nu}\nabla_{\nu}\hat{e}_{(b)\mu},
\end{align} 

\noindent so the directional derivative finally takes the form

\begin{align}
     A_{(a),(b)}=\hat{e}_{(a)}\!^{\nu}\hat{e}_{(b)}\!^{\mu}\nabla_{\mu}A_{\nu} + \gamma_{(c)(a)(b)}A^{(c)}.
\end{align}

The fact that the matrix $\eta_{(a)(b)}$ is constant implies that the Ricci rotation coefficients are antisymmetric in the first two indices \cite{chandrasekhar1998mathematical}

\begin{align}
    \gamma_{(a)(b)(c)} = -\gamma_{(b)(a)(c)}.
\end{align}

We can also define the \textit{intrinsic derivative} as \cite{alcubierre2008introduction}

\begin{eqnarray}\label{ec:intr}
    A_{(a)|(b)} &=& \hat{e}_{(a)}\!^{\mu}\hat{e}_{(b)}\!^{\nu}\nabla_{\nu}A_{\mu} \nonumber \\
                &=& A_{(a),(b)}-\gamma^{(c)}_{\quad(a)(b)}A_{(c)},
\end{eqnarray}

\noindent where $\gamma^{(c)}_{\quad(a)(b)}=\eta^{(c)(d)}\gamma_{(d)(a)(b)}$. The intrinsic derivative corresponds to the covariant derivative expressed in the tetrad basis, and the Ricci rotation coefficients correspond to the connection coefficients in the tetrad basis. Eqs.~(\ref{ec:intr}) generalizes analogously for a tensor of arbitrary rank

\begin{eqnarray}
    T_{(a_{1})\dots(a_{n})|(b)} &=& T_{(a_{1})\dots(a_{n}),(b)}-\gamma^{(c)}_{\quad(a_{1})(b)}T_{(c)\dots(a_{n})}  \nonumber  \\
                                &-& \dots-\gamma^{(c)}_{\quad(a_{n})(b)}T_{(a_{1})\dots(c)}.
\end{eqnarray}

The commutation of directional derivatives along the tetrads is given by \cite{chandrasekhar1998mathematical}

\begin{align}\label{eq:comt}
    [\hat{e}_{(a)}\!^{\mu}\nabla_{\mu},\hat{e}_{(b)}\!^{\nu}\nabla_{\nu}]=C^{(c)}_{\quad(a)(b)}\hat{e}_{(c)}\!^{\lambda}\nabla_{\lambda}.
\end{align}

\noindent where $C^{(c)}_{\quad(a)(b)}$ are the \textit{structure constants}, which can be obtained from the Ricci rotation coefficients via

\begin{align}
    C^{(c)}_{\quad(a)(b)} = \gamma^{(c)}_{\quad(b)(a)}-\gamma^{(c)}_{\quad(a)(b)}.
\end{align}

Using the tetrad vectors in the definition of the Riemann tensor \cite{wald2010general}, we can find an expression for this tensor in terms of the Ricci rotation coefficients \cite{alcubierre2008introduction}

\begin{eqnarray}\label{ec:rimT}
    R_{(a)(b)(m)(n)}  &=& \gamma_{(a)(b)(n),(m)} - \gamma_{(a)(b)(m),(n)} \nonumber \\
                      &+& \gamma_{(a)(b)(c)}\left(\gamma^{(c)}_{\quad(m)(n)} -\gamma^{(c)}_{\quad(n)(m)}\right)  \\
                      &+& \gamma_{(a)(c)(m)}\gamma^{(c)}_{\quad(b)(n)}-\gamma_{(a)(c)(n)}\gamma^{(c)}_{\quad(b)(m)}. \nonumber
\end{eqnarray}

\noindent where

\begin{align}
    R_{(a)(b)(m)(n)} = R_{\alpha\beta\mu\nu}\hat{e}_{(a)}\!^{\alpha}\hat{e}_{(b)}\!^{\beta}\hat{e}_{(m)}\!^{\mu}\hat{e}_{(n)}\!^{\nu}.
\end{align}

\noindent corresponds to the projection of the Riemann tensor onto the tetrad basis.

Similarly, the Bianchi identity (Eq.~(\ref{ec:bianchi}))

\begin{eqnarray}\label{ec:bianchi}
    \nabla_\lambda R_{\alpha \beta \mu \nu} + \nabla_\beta R_{ \lambda \alpha \mu \nu} + \nabla_\alpha R_{ \beta \lambda \mu \nu} = 0.
\end{eqnarray}

\noindent in the tetrad basis takes the form \cite{chandrasekhar1998mathematical}

\begin{eqnarray}\label{eq:biT}
    R_{(a)(b)(c)(d)|(f)} + R_{(a)(b)(f)(c)|(d)} + R_{(a)(b)(d)(f)|(c)} = 0.
\end{eqnarray}

Eqs. (\ref{eq:comt}), (\ref{ec:rimT}) and (\ref{eq:biT}) will be the basic relations through which curvature is described in the tetrad formalism.

\subsection{Newman-Penrose Formalism}

Starting from the tetrad formalism, the NP formalism is constructed by choosing a particular complex tetrad basis, which will be known as the \textit{null tetrad}, and from which all the quantities describing the curvature of spacetime will be built.

\subsubsection{Null tetrads}

The null tetrad will be given by the vectors $(l,n,m,\bar{m})$, which are identified with the tetrad basis vectors as

\begin{align}
    (l^{\mu},n^{\mu},m^{\mu},\bar{m}^{\mu})=(\hat{e}_{(0)}^{\mu},\hat{e}_{(1)}^{\mu},\hat{e}_{(2)}^{\mu},\hat{e}_{(3)}^{\mu}).
\end{align}

The vectors $l$ and $n$ are real, while the vectors $m$ and $\bar{m}$ are complex, being complex conjugates of each other \cite{chandrasekhar1998mathematical}. Likewise, these vectors are null and satisfy the following conditions \cite{alcubierre2008introduction}

\begin{eqnarray}\label{eq:normalizatetrada}
    l^{\mu}l_{\mu}=n^{\mu}n_{\mu}=m^{\mu}m_{\mu}=\bar{m}^{\mu}\bar{m}_{\mu}=0,   \nonumber  \\
    l^{\mu}m_{\mu}=l^{\mu}\bar{m}_{\mu}=n^{\mu}m_{\mu}=n^{\mu}\bar{m}_{\mu}=0,\\
    l^{\mu}n_{\mu}=-m^{\mu}\bar{m}_{\mu}=-1.      \nonumber  
\end{eqnarray}

From these conditions, we observe that the inverse basis is given by

\begin{align}
    (\hat{e}^{(0)}_{\ \mu},\hat{e}^{(1)}_{\ \mu},\hat{e}^{(2)}_{\ \mu},\hat{e}^{(3)}_{\ \mu})=(-n_{\mu},-l_{\mu},\bar{m}_{\mu},m_{\mu}),
\end{align} 

\noindent and the components of the matrix $\eta_{(a)(b)}$ and its inverse $\eta^{(a)(b)}$ take the following values \cite{frolov2012black}

\begin{align}
  \eta_{(a)(b)}= \eta^{(a)(b)}= \left(\begin{array}{cccc}
        0 & -1 & 0 & 0 \\
        -1 & 0 & 0 & 0 \\
        0 & 0 & 0 & 1 \\
        0 & 0 & 1 & 0  
    \end{array}\right)
\end{align}

Using Eq.~(\ref{eq:tcomp}), we can write the metric in terms of the null tetrad, obtaining \cite{alcubierre2008introduction}

\begin{align}\label{eq:metricatetradas}
    g_{\mu\nu}=-l_{\mu}n_{\nu}-n_{\mu}l_{\nu}+m_{\mu}\bar{m}_{\nu}+\bar{m}_{\mu}m_{\nu}.
\end{align}

\subsubsection{Newman-Penrose Quantities}

In the NP formalism, the different quantities constructed from the tetrad and the projections of the curvature tensors onto this basis take special names and symbols.

The covariant directional derivatives along the tetrad directions are denoted by \cite{newman1962approach}

\begin{eqnarray}
    D &=& l^{\mu}\nabla_{\mu}\ , \quad \Delta=n^{\mu}\nabla_{\mu}, \nonumber \\ 
    \delta &=& m^{\mu}\nabla_{\mu}\ , \quad \bar{\delta}=\bar{m}^{\mu}\nabla_{\mu}\ .
\end{eqnarray} 

The Ricci rotation coefficients are represented by twelve complex scalar quantities, known as the \textit{spin coefficients} \cite{frolov2012black}

\begin{eqnarray}\label{eq definicion coeficientes de spin}
    -\kappa &=& \gamma_{(2)(0)(0)}=m^{\mu}Dl_{\mu}\ ,\nonumber \\
    -\rho &=&\gamma_{(2)(0)(3)}=m^{\mu}\bar{\delta}l_{\mu}\ , \nonumber \\
    -\sigma &=&\gamma_{(2)(0)(2)}=m^{\mu}\delta l_{\mu}\ ,\nonumber \\
    -\tau &=& \gamma_{(2)(0)(1)}=m^{\mu}\Delta l_{\mu}\ , \nonumber  \\
     \nu &=& \gamma_{(3)(1)(1)}=\bar{m}^{\mu}\Delta n_{\mu}\ ,\nonumber \\
     \mu &=& \gamma_{(3)(1)(2)}=\bar{m}^{\mu}\delta n_{\mu}\ , \nonumber  \\
     \lambda &=& \gamma_{(3)(1)(3)}=\bar{m}^{\mu}\bar{\delta} n_{\mu}\ ,\nonumber \\
     \pi &=& \gamma_{(3)(1)(0)}=\bar{m}^{\mu}D n_{\mu}\ , \nonumber  \\
     -\varepsilon &=& \dfrac{1}{2}(\gamma_{(1)(0)(0)}-\gamma_{(3)(2)(0)}) \nonumber \\
     &=& \dfrac{1}{2}(n^{\mu}Dl_{\mu}-\bar{m}^{\mu}Dm_{\mu})\ ,  \\
     -\beta &=& \dfrac{1}{2}(\gamma_{(1)(0)(2)}-\gamma_{(3)(2)(2)}) \nonumber \\ 
     &=& \dfrac{1}{2}(n^{\mu}\delta l_{\mu}-\bar{m}^{\mu}\delta m_{\mu})\ , \nonumber \\
     -\gamma &=& \dfrac{1}{2}(\gamma_{(2)(3)(1)}-\gamma_{(0)(1)(1)}) \nonumber \\
     &=& \dfrac{1}{2}(n^{\mu}\Delta l_{\mu}-\bar{m}^{\mu}\Delta m_{\mu})\ , \nonumber \\
     -\alpha &=& \dfrac{1}{2}(\gamma_{(2)(3)(3)}-\gamma_{(0)(1)(3)}) \nonumber \\ 
     &=& \dfrac{1}{2}(n^{\mu}\bar{\delta}l_{\mu}-\bar{m}^{\mu}\bar{\delta}m_{\mu})\ . \nonumber 
\end{eqnarray}

The components of the Ricci tensor are represented by three complex scalars

\begin{align}
    \begin{aligned}
   & \Phi_{01}=\bar{\Phi}_{10}=\dfrac{1}{2}R_{(0)(2)}=\dfrac{1}{2}R_{\mu\nu}l^{\mu}m^{\nu}\ ,\\
&    \Phi_{02}=\bar{\Phi}_{20}=\dfrac{1}{2}R_{(2)(2)}=\dfrac{1}{2}R_{\mu\nu}m^{\mu}m^{\nu}\ ,\\
 &   \Phi_{12}=\bar{\Phi}_{21}=\dfrac{1}{2}R_{(1)(2)}=\dfrac{1}{2}R_{\mu\nu}n^{\mu}m^{\nu}\ .
    \end{aligned}
\end{align} 

\noindent and four real scalars

\begin{eqnarray}
     \Phi_{00} &=& \dfrac{1}{2}R_{(0)(0)} = \dfrac{1}{2}R_{\mu\nu}l^{\mu}l^{\nu}\ ,   \nonumber\\
     \Phi_{22} &=& \dfrac{1}{2}R_{(1)(1)} = \dfrac{1}{2}R_{\mu\nu}n^{\mu}n^{\nu}\ ,\nonumber\\
     \Phi_{11} &=& \dfrac{1}{4}(R_{(0)(1)} + R_{(2)(3)})  \\
               &=&\dfrac{1}{4}R_{\mu\nu}(l^{\mu}n^{\nu}+m^{\mu}\bar{m}^{\nu})\ ,\nonumber\\
     \Lambda   &=& \dfrac{1}{24}R = \dfrac{1}{12}(R_{(2)(3)}-R_{(0)(1)}) \nonumber \nonumber\\
               &=& \dfrac{1}{12}R_{\mu\nu}(m^{\mu}\bar{m}^{\nu}-l^{\mu}n^{\nu})\ . \nonumber
\end{eqnarray}

\noindent known as the \textit{Ricci scalars} \cite{alcubierre2008introduction}.

Finally, five complex scalar quantities known as the \textit{Weyl scalars} are defined, representing the components of the Weyl tensor \cite{alcubierre2008introduction}

\begin{align}\label{eq:weylscaltet}
        \begin{aligned}
        & \Psi_{0}=C_{(0)(2)(0)(2)}= C_{\alpha\beta\mu\nu}l^{\alpha}m^{\beta}l^{\mu}m^{\nu}\ ,\\
        & \Psi_{1}=C_{(0)(1)(0)(2)}= C_{\alpha\beta\mu\nu}l^{\alpha}n^{\beta}l^{\mu}m^{\nu}\ ,\\
        & \Psi_{2}=C_{(0)(2)(3)(1)}= C_{\alpha\beta\mu\nu}l^{\alpha}m^{\beta}\bar{m}^{\mu}n^{\nu}\ ,\\
        & \Psi_{3}=C_{(0)(1)(3)(1)}= C_{\alpha\beta\mu\nu}l^{\alpha}n^{\beta}\bar{m}^{\mu}n^{\nu}\ ,\\
        & \Psi_{4}=C_{(1)(3)(1)(3)}= C_{\alpha\beta\mu\nu}n^{\alpha}\bar{m}^{\beta}n^{\mu}\bar{m}^{\nu}\ .\\
    \end{aligned}
\end{align}

From these quantities, the equations that describe the curvature of spacetime will be constructed. Likewise, from the definitions of the NP quantities, we observe that if we perform the exchange

\begin{align}\label{eq:cambio tetradas}
    l^{\mu}\longleftrightarrow n^{\mu}\ , \quad m^{\mu}\longleftrightarrow \bar{m}^{\mu}\ ,
\end{align} 

\noindent they transform as follows \cite{geroch1973space}

\begin{subequations} \label{cambiotodo}
\begin{align}
    D\longleftrightarrow\Delta\ , \quad \delta\longleftrightarrow\bar{\delta}\ ,
\end{align}

\begin{align}
    \Psi_{0}\longleftrightarrow\Psi_{4}\ , \quad \Psi_{2}\rightarrow\Psi_{2}\ , \quad \Psi_{1}\longleftrightarrow\Psi_{3}\ ,
\end{align}

\begin{align}
    \begin{aligned}
    \kappa\longleftrightarrow-\nu\ , \quad \sigma\longleftrightarrow-\lambda\ , \quad \rho\longleftrightarrow-\mu\ ,\\
    \tau\longleftrightarrow-\pi\ , \quad \beta\longleftrightarrow-\alpha\ , \quad \varepsilon\longleftrightarrow-\gamma\ ,
    \end{aligned}
\end{align}

\begin{align}
    \begin{aligned}
    \Phi_{01}\longleftrightarrow\Phi_{21}\ & , \quad \Phi_{02}\longleftrightarrow\Phi_{20}\ , \quad \Phi_{00}\longleftrightarrow\Phi_{22}\\
    & \Phi_{11}\rightarrow\Phi_{11}\ , \quad \Lambda\rightarrow\Lambda\ .
    \end{aligned}
\end{align}
\end{subequations}

It is important to emphasize that the sign conventions in the definitions of these quantities are used for the signature $-+++$, as is the case in the references \cite{alcubierre2008introduction,frolov2012black,griffiths2009exact}. When using the signature $+---$, some quantities will have different signs, which can be consulted in references \cite{chandrasekhar1998mathematical, newman1962approach}. Despite these differences, the conventions are chosen so that the field equations presented in the next section remain invariant.

\subsubsection{Newman-Penrose Field Equations}

As previously indicated, the commutation relations, the definition of the Riemann tensor, and the Bianchi identity provide us with the description of spacetime curvature in the tetrad formalism. In the case of the NP formalism, we can write these equations in terms of the quantities defined in the previous section.

The commutation relations (Eq.~\ref{eq:comt}) take the following form \cite{o2003introduction}

\begin{eqnarray}\label{eq:relconmNP}
    \Delta D - D \Delta &=& (\gamma + \bar{\gamma}) D + (\varepsilon + \bar{\varepsilon}) \Delta  \nonumber    \nonumber  \\
                        &-& (\bar{\tau} + \pi) \delta  - (\tau + \bar{\pi}) \bar{\delta} \ ,    \nonumber  \\
    \delta D - D \delta &=& (\bar{\alpha} + \beta - \bar{\pi}) D + \kappa \Delta \nonumber \\
                        &-& (\bar{\rho} + \varepsilon - \bar{\varepsilon}) \delta - \sigma \bar{\delta} \ ,    \nonumber  \\
    \delta \Delta - \Delta \delta &=& -\bar{\nu} D + (\tau - \bar{\alpha} - \beta) \Delta  \\
                        &+& (\mu - \gamma + \bar{\gamma}) \delta + \bar{\lambda} \bar{\delta} \ ,    \nonumber  \\
    \bar{\delta} \delta - \delta \bar{\delta} &=& (\bar{\mu} - \mu) D + (\bar{\rho} - \rho) \Delta \nonumber \\
                        &+& (\alpha - \bar{\beta}) \delta + (\beta - \bar{\alpha}) \bar{\delta} \ .   \nonumber  
\end{eqnarray}

In contrast to Einstein's field equations, which are second order in the derivatives of the metric $g_{\mu\nu}$, the field equations in the NP formalism are first order in the derivatives of the quantities. Moreover, the complete set of equations remains invariant under the transformation (\ref{eq:cambio tetradas}) \cite{geroch1973space}.

From NP equations, it can directly solve for the following expressions for the Weyl scalars

\begin{eqnarray}  \label{eq:definicion weyl}
   \Psi_{0}  &=& D\sigma - \delta \kappa - \sigma(\rho + \bar{\rho} + 3 \varepsilon - \bar{\varepsilon}) \nonumber \\
             &+& \kappa(\tau - \bar{\pi} + \bar{\alpha} + 3 \beta) \ ,   \nonumber    \\
    \Psi_{1} &=& D\beta - \delta \varepsilon - \sigma(\alpha + \pi) - \beta(\bar{\rho} - \bar{\varepsilon}) \nonumber \\
             &+& \kappa(\mu + \gamma) + \varepsilon(\bar{\alpha} - \bar{\pi}) \ ,   \nonumber    \\
    \Psi_{2} &=& \bar{\delta} \tau - \Delta \rho - (\rho \bar{\nu} + \sigma \lambda) + \tau(\bar{\beta} - \alpha - \bar{\tau}) \nonumber \\
             &+& \rho(\gamma + \bar{\gamma}) + \nu \kappa - 2 \Lambda \ ,\\
    \Psi_{3} &=& \bar{\delta} \gamma - \Delta \alpha + \nu(\rho + \varepsilon) - \lambda(\tau + \beta) \nonumber \\
             &+& \alpha(\bar{\gamma} - \bar{\mu}) + \gamma(\bar{\beta} - \bar{\tau}) \ ,    \nonumber     \\ 
    \Psi_{4} &=& \bar{\delta} \nu - \Delta \lambda - \lambda(\mu + \bar{\mu} + 3 \gamma - \bar{\gamma})  \nonumber \\
             &+& \nu(3 \alpha + \bar{\beta} + \pi - \bar{\tau}) \ .    \nonumber    
\end{eqnarray}

\noindent and the Ricci scalars

\begin{eqnarray}  \label{eq:definicion Ricci}
    \Phi_{00} &=& D \rho - \bar{\delta} \kappa - (\rho^{2} + \sigma \bar{\sigma}) - \rho (\varepsilon + \bar{\varepsilon}) \nonumber \\
              &+& \bar{\kappa} \tau + \kappa (3 \alpha + \bar{\beta} - \pi) \ ,  \nonumber   \\
    \Phi_{10} &=& D \alpha - \bar{\delta} \varepsilon - \alpha (\rho + \bar{\varepsilon} - 2 \varepsilon) - \beta \bar{\sigma} \nonumber    \nonumber    \\
              &+& \bar{\beta} \varepsilon + \kappa \lambda + \bar{\kappa} \gamma - \pi (\varepsilon + \rho) \ ,    \nonumber    \\
    \Phi_{20} &=& D \lambda - \bar{\delta} \pi - (\rho \lambda + \bar{\sigma} \mu) - \pi (\pi + \alpha - \beta)  \nonumber \\
              &+& \nu \bar{\kappa} + \lambda (3 \varepsilon - \bar{\varepsilon}) \ ,   \nonumber     \\
    \Phi_{12} &=& \delta \gamma - \Delta \beta - \gamma (\tau - \bar{\alpha} - \beta) - \mu \tau + \sigma \nu   \nonumber \\
              &+& \varepsilon \bar{\nu} + \beta (\gamma - \bar{\gamma} - \mu) - \alpha \bar{\lambda} \ , \\
    \Phi_{22} &=& \delta \nu - \Delta \mu - (\mu^{2} + \lambda \bar{\lambda}) - \mu (\gamma + \bar{\gamma})  \nonumber \\
              &+& \bar{\nu} \pi - \nu (\tau - 3 \beta - \bar{\alpha}) \ ,   \nonumber   \\
    2 \Phi_{11} &=& D \gamma - \Delta \varepsilon - \alpha (\tau + \bar{\pi}) - \beta (\bar{\tau} + \pi)  \nonumber \\
              &+& \gamma (\varepsilon + \bar{\varepsilon}) + \varepsilon (\gamma + \bar{\gamma}) - \tau \pi + \nu \kappa  \nonumber \\
              &+& \delta \alpha - \bar{\delta} \beta - (\mu \rho - \lambda \sigma) - \alpha \bar{\alpha} - \beta \bar{\beta}   \nonumber \\
              &+& 2 \alpha \beta - \gamma (\rho - \bar{\rho}) - \varepsilon (\mu - \bar{\mu}) \ .     \nonumber   
\end{eqnarray}

If the null tetrad is known, from which the metric, spin coefficients, and curvature scalar are obtained, one can use the above expressions to compute the Weyl and Ricci scalars without needing the components of the Riemann, Weyl, and Ricci tensors beforehand.

Using Einstein's field equations, the Ricci scalars can be related to the components of the energy-momentum tensor by \cite{teukolsky1973perturbations}

\begin{eqnarray}\label{eq:NPTEI}
    \Phi_{01} &=& 4 \pi T_{(0)(2)} = 4 \pi T_{\mu \nu} l^{\mu} m^{\nu}, \nonumber \\
    \Phi_{02} &=& 4 \pi T_{(2)(2)} = 4 \pi T_{\mu \nu} m^{\mu} m^{\nu}, \nonumber \\
    \Phi_{12} &=& 4 \pi T_{(1)(2)} = 4 \pi T_{\mu \nu} n^{\mu} m^{\nu}, \nonumber \\
    \Phi_{00} &=& 4 \pi T_{(0)(0)} = 4 \pi T_{\mu \nu} l^{\mu} l^{\nu}, \nonumber \\
    \Phi_{22} &=& 4 \pi T_{(1)(1)} = 4 \pi T_{\mu \nu} n^{\mu} n^{\nu}, \\
    \Phi_{11} &=& 2 \pi (T_{(0)(1)} + T_{(2)(3)})   \nonumber \\
              &=& 2 \pi T_{\mu \nu} (l^{\mu} n^{\nu} + m^{\mu} \bar{m}^{\nu}), \nonumber \\
    \Lambda   &=& - \dfrac{\pi}{3} T = - \dfrac{2 \pi}{3} (T_{(2)(3)} - T_{(0)(1)})  \nonumber \\
              &=& - \dfrac{2 \pi}{3} T_{\mu \nu} (m^{\mu} \bar{m}^{\nu} - l^{\mu} n^{\nu}). \nonumber
\end{eqnarray}

\noindent where $\pi$ is the constant and not the spin coefficient. Both the constant and the spin coefficient are denoted by the letter $\pi$, however their difference in the context of the equations will be evident, so there should be no confusion. In vacuum, the Ricci scalars vanish, so the NP field equations reduce considerably \cite{chandrasekhar1998mathematical}.

\subsubsection{Scalar and Electromagnetic Fields in the Newman-Penrose Formalism}

Scalar and electromagnetic fields can also be described using the NP formalism. In the case of the electromagnetic field, the components of the field tensor $F_{\mu\nu}$ are represented by three complex quantities known as the \textit{Maxwell scalars} \cite{chandrasekhar1998mathematical}

\begin{eqnarray}
       \phi_{0} &=& F_{(0)(2)} = F_{\mu\nu} l^{\mu} m^{\nu} , \nonumber  \\
       \phi_{1} &=& \dfrac{1}{2} (F_{(0)(1)} + F_{(3)(2)}) \nonumber \\
                &=& \dfrac{1}{2} F_{\mu\nu} (l^{\mu} n^{\nu} + \bar{m}^{\mu} m^{\nu}) \ , \\
       \phi_{2} &=& F_{(4)(2)} = F_{\mu\nu} \bar{m}^{\mu} m^{\nu}  .  \nonumber
\end{eqnarray}

These transform under the change (\ref{eq:cambio tetradas}) as

\begin{align}\label{cambioMAX}
    \phi_{0} \longleftrightarrow -\phi_{2} \ , \quad \phi_{1} \longrightarrow -\phi_{1} \ .
\end{align}

Projecting Maxwell’s equations (Eq.~\ref{MQE}) onto the null tetrad and expressing them in terms of intrinsic derivatives, for vacuum ($J^{\mu} = 0$), we obtain the following four complex equations \cite{chandrasekhar1998mathematical}

\begin{eqnarray}\label{eq:MAXNPASFASDF}
    D \phi_{1} - \bar{\delta} \phi_{0} &=& (\pi - 2 \alpha) \phi_{0} + 2 \rho \phi_{1} - \kappa \phi_{2} \ ,  \nonumber \\
    D \phi_{2} - \bar{\delta} \phi_{1} &=& - \lambda \phi_{0} + 2 \pi \phi_{1} + (\rho - 2 \varepsilon) \phi_{2} \ , \\
    \delta \phi_{1} - \Delta \phi_{0} &=& (\mu - 2 \gamma) \phi_{0} + 2 \tau \phi_{1} - \sigma \phi_{2} \ ,  \nonumber     \\
    \delta \phi_{2} - \Delta \phi_{1} &=& - \nu \phi_{0} + 2 \mu \phi_{1} + (\tau - 2 \beta) \phi_{2} \ .    \nonumber 
\end{eqnarray}

The components of the electromagnetic energy-momentum tensor (Eq.~\ref{eq:Tmax}), in the tetrad basis, are given by \cite{griffiths2009exact}

\begin{eqnarray}\label{eq:tensorenergiamax}
    T_{(0)(0)} &=& \dfrac{1}{2\pi} \phi_{0} \bar{\phi}_{0} , \nonumber \\ 
    T_{(0)(2)} &=& \dfrac{1}{2\pi} \phi_{0} \bar{\phi}_{1} , \nonumber \\
    T_{(1)(1)} &=& \dfrac{1}{2\pi} \phi_{2} \bar{\phi}_{2} , \nonumber \\
    T_{(2)(2)} &=& \dfrac{1}{2\pi} \phi_{0} \bar{\phi}_{2} , \\
    T_{(1)(2)} &=& \dfrac{1}{2\pi} \phi_{1} \bar{\phi}_{2} , \nonumber \\
    T_{(0)(1)} + T_{(2)(3)} &=& \dfrac{1}{\pi} \phi_{1} \bar{\phi}_{1} , \nonumber \\
    T = 2 (T_{(2)(3)} &-& T_{(0)(1)}) = 0  . \nonumber
\end{eqnarray}

In the case of a massless scalar field $\Phi$, writing the covariant derivatives in terms of the directional derivatives of the null tetrad, we find the following expression for the Klein-Gordon equation (Eq.~\ref{KGEQ}) in terms of NP quantities \cite{silvaortigoza1996solution}

\begin{align}\label{KGNPSADF}
    \begin{aligned}
  & \left[(D + \varepsilon + \bar{\varepsilon} - \bar{\rho} - \rho) \Delta - (\delta + \beta - \bar{\alpha} + \bar{\pi} - \tau) \bar{\delta} \right. \\
  & \left. + (\Delta - \gamma - \bar{\gamma} + \mu + \bar{\mu}) D - (\bar{\delta} - \alpha + \bar{\beta} - \bar{\tau} + \pi) \delta \right] \Phi = 0 \ .
    \end{aligned}
\end{align}

Also, the components of the energy-momentum tensor (Eq.~\ref{eq:Tklein}) in the tetrad basis are given by

\begin{eqnarray}
    T_{(0)(0)} &=& (D \Phi)^{2}, \nonumber \\
    T_{(0)(2)} &=& D \Phi \, \delta \Phi ,  \nonumber \\
    T_{(1)(1)} &=& (\Delta \Phi)^{2}  , \nonumber \\ 
    T_{(2)(2)} &=& (\delta \Phi)^{2} ,  \\
    T_{(1)(2)} &=& \Delta \Phi \, \delta \Phi , \nonumber  \\
    T_{(0)(1)} &+& T_{(2)(3)} = D \Phi \, \Delta \Phi + \delta \Phi \, \bar{\delta} \Phi , \nonumber \\
    T = 2 (T_{(2)(3)} &-& T_{(0)(1)}) = 2 (D \Phi \, \Delta \Phi - \delta \Phi \, \bar{\delta} \Phi)  .  \nonumber
\end{eqnarray}

Through Eq.~(\ref{eq:NPTEI}), the Ricci scalars are related to the components of the above energy-momentum tensors, thus completing the coupling between scalar and electromagnetic fields and the curvature of spacetime.


Using the definition of the Riemann tensor in terms of the Ricci rotation coefficients (Eq.~\ref{ec:rimT}), together with the definition of the Weyl tensor (Eq. \ref{eq:weyl})

\begin{eqnarray}\label{eq:weyl}
    C_{\alpha \beta \mu \nu} &=& R_{\alpha \beta \mu \nu} + \frac{1}{(n-1)(n-2)} \left(  g_{\alpha \mu} g_{\nu \beta} - g_{\alpha \nu} g_{\mu \beta}  \right) R \nonumber \\
    &-& \frac{1}{(n-2)} \left(  g_{\alpha \mu} R_{\nu \beta} -  g_{\alpha \nu} R_{\mu \beta}  + g_{\beta \nu} R_{\mu \alpha} - g_{\beta \mu} R_{\nu \alpha}  \right).
\end{eqnarray}

\noindent projected on the null tetrad, one obtains a set of 18 complex equations known as the \textit{Newman-Penrose equations} \cite{chandrasekhar1998mathematical}

\begin{eqnarray}\label{ec:NPE}
    D \rho - \bar{\delta} \kappa = (\rho^{2} + \sigma \bar{\sigma}) + \rho (\varepsilon + \bar{\varepsilon}) - \bar{\kappa} \tau - \kappa (3 \alpha + \bar{\beta} - \pi) + \Phi_{00} \ ,   \\
    D \sigma - \delta \kappa = \sigma (\rho + \bar{\rho} + 3 \varepsilon - \bar{\varepsilon}) - \kappa (\tau - \bar{\pi} + \bar{\alpha} + 3 \beta) + \Psi_{0} \ , \label{eq:NP2}      \\ 
    D \tau - \Delta \kappa = \rho (\tau + \bar{\pi}) + \sigma (\bar{\tau} + \pi) + \tau (\varepsilon - \bar{\varepsilon}) - \kappa (3 \gamma + \bar{\gamma}) + \Psi_{1} + \Phi_{01} \ , \label{eq:NP3}\\
    D \alpha - \bar{\delta} \varepsilon = \alpha (\rho + \bar{\varepsilon} - 2 \varepsilon) + \beta \bar{\sigma} - \bar{\beta} \varepsilon - \kappa \lambda - \bar{\kappa} \gamma + \pi (\varepsilon + \rho) + \Phi_{10} \ ,\\
    D \beta - \delta \varepsilon = \sigma (\alpha + \pi) + \beta (\bar{\rho} - \bar{\varepsilon}) - \kappa (\mu + \gamma) - \varepsilon (\bar{\alpha} - \bar{\pi}) + \Psi_{1} \ , \label{eq:NP5}\\
    D \gamma - \Delta \varepsilon = \alpha (\tau + \bar{\pi}) + \beta (\bar{\tau} + \pi) - \gamma (\varepsilon + \bar{\varepsilon}) - \varepsilon (\gamma + \bar{\gamma}) + \tau \pi - \nu \kappa + \Psi_{2} + \Phi_{11} - \Lambda \ , \\ 
    D \lambda - \bar{\delta} \pi = (\rho \lambda + \bar{\sigma} \mu) + \pi (\pi + \alpha - \beta) - \nu \bar{\kappa} - \lambda (3 \varepsilon - \bar{\varepsilon}) + \Phi_{20} \ ,\\
    D \mu - \delta \pi = (\bar{\rho} \mu + \sigma \lambda) + \pi (\bar{\pi} - \bar{\alpha} + \beta) - \mu (\varepsilon + \bar{\varepsilon}) - \nu \kappa + \Psi_{2} + 2 \Lambda \ ,\\
    D \nu - \Delta \pi = \mu (\pi + \bar{\tau}) + \lambda (\bar{\pi} + \tau) + \pi (\gamma - \bar{\gamma}) - \nu (3 \varepsilon + \bar{\varepsilon}) + \Psi_{3} + \Phi_{21} \ , \\ 
    \Delta \lambda - \bar{\delta} \nu = - \lambda (\mu + \bar{\mu} + 3 \gamma - \bar{\gamma}) + \nu (3 \alpha + \bar{\beta} + \pi - \bar{\tau}) - \Psi_{4} \ , \label{eq:NP10} \\
    \delta \rho - \bar{\delta} \sigma = \rho (\bar{\alpha} + \beta) - \sigma (3 \alpha - \bar{\beta}) + \tau (\rho - \bar{\rho}) + \kappa (\mu - \bar{\mu}) - \Psi_{1} + \Phi_{01} \ , \label{eq:NP11}\\
    \delta \alpha - \bar{\delta} \beta = (\mu \rho - \lambda \sigma) + \alpha \bar{\alpha} + \beta \bar{\beta} - 2 \alpha \beta + \gamma (\rho - \bar{\rho}) + \varepsilon (\mu - \bar{\mu}) - \Psi_{2} + \Phi_{11} + \Lambda \ , \\
    \delta \lambda - \bar{\delta} \mu = \nu (\rho - \bar{\rho}) + \pi (\mu - \bar{\mu}) + \mu (\alpha + \bar{\beta}) + \lambda (\bar{\alpha} - 3 \beta) - \Psi_{3} + \Phi_{21} \ , \\
    \delta \nu - \Delta \mu = (\mu^{2} + \lambda \bar{\lambda}) + \mu (\gamma + \bar{\gamma}) - \bar{\nu} \pi + \nu (\tau - 3 \beta - \bar{\alpha}) + \Phi_{22} \ , \\
    \delta \gamma - \Delta \beta = \gamma (\tau - \bar{\alpha} - \beta) + \mu \tau - \sigma \nu - \varepsilon \bar{\nu} - \beta (\gamma - \bar{\gamma} - \mu) + \alpha \bar{\lambda} + \Phi_{12} \ , \\
    \delta \tau - \Delta \sigma = (\mu \sigma + \bar{\lambda} \rho) + \tau (\tau + \beta - \bar{\alpha}) - \sigma (3 \gamma - \bar{\gamma}) - \kappa \bar{\nu} + \Phi_{02} \ , \\
    \Delta \rho - \bar{\delta} \tau = -(\rho \bar{\nu} + \sigma \lambda) + \tau (\bar{\beta} - \alpha - \bar{\tau}) + \rho (\gamma + \bar{\gamma}) + \nu \kappa - \Psi_{2} - 2 \Lambda \ , \\
    \Delta \alpha - \bar{\delta} \gamma = \nu (\rho + \varepsilon) - \lambda (\tau + \beta) + \alpha (\bar{\gamma} - \bar{\mu}) + \gamma (\bar{\beta} - \bar{\tau}) - \Psi_{3} \ .
\end{eqnarray}

The Bianchi identity (Eq.~\ref{eq:biT}) in terms of the NP quantities takes the following form \cite{frolov2012black}

\begin{subequations}
\begin{align}\label{eq:bianchi1}
    \begin{aligned}
    \bar{\delta}\Psi_{0} - D\Psi_{1} + D\Phi_{01} - \delta\Phi_{00} = (4\alpha - \pi)\Psi_{0} - 2(2\rho + \varepsilon)\Psi_{1} + 3\kappa \Psi_{2}\\
    + (\bar{\pi} - 2\bar{\alpha} - 2\beta)\Phi_{00} + 2(\varepsilon + \bar{\rho})\Phi_{01} + 2\sigma \Phi_{10} - 2\kappa \Phi_{11} - \bar{\kappa} \Phi_{02} \ ,
    \end{aligned}
 \end{align}

\begin{align}\label{eq:bianchi2}
    \begin{aligned}
    \Delta \Psi_{0} - \delta \Psi_{1} + D \Phi_{02} - \delta \Phi_{01} = (4\gamma - \mu) \Psi_{0} - 2(2\tau + \beta) \Psi_{1} + 3 \sigma \Psi_{2}\\
    + (2\varepsilon - 2\bar{\varepsilon} + \bar{\rho}) \Phi_{02} + 2(\bar{\pi} - \beta) \Phi_{01} + 2 \sigma \Phi_{11} - 2 \kappa \Phi_{12} - \bar{\lambda} \Phi_{00} \ ,
    \end{aligned}
 \end{align}

\begin{align}\label{eq:bianchi3}
    \begin{aligned}
    \bar{\delta} \Psi_{3} - D \Psi_{4} + \bar{\delta} \Phi_{21} - \Delta \Phi_{20} = (4\varepsilon - \rho) \Psi_{4} - 2(2\pi + \alpha) \Psi_{3} + 3 \lambda \Psi_{2}\\
    + (2\gamma - 2\bar{\gamma} + \bar{\mu}) \Phi_{20} + 2(\bar{\tau} - \alpha) \Phi_{21} + 2 \lambda \Phi_{11} - 2 \nu \Phi_{10} - \bar{\sigma} \Phi_{22} \ ,
    \end{aligned}
 \end{align}

\begin{align}\label{eq:bianchi4}
    \begin{aligned}
    \Delta \Psi_{3} - \delta \Psi_{4} + \bar{\delta} \Phi_{22} - \Delta \Phi_{21} = (4\beta - \tau) \Psi_{4} - 2(2\mu + \gamma) \Psi_{3} + 3 \nu \Psi_{2}\\
    + (\bar{\tau} - 2\bar{\beta} - 2 \alpha) \Phi_{22} + 2(\gamma + \bar{\mu}) \Phi_{21} + 2 \lambda \Phi_{12} - 2 \nu \Phi_{11} - \bar{\nu} \Phi_{20} \ ,
    \end{aligned}
 \end{align}

 \begin{align}\label{eq:bianchi5}
    \begin{aligned}
    D \Psi_{2} - \bar{\delta} \Psi_{1} + \Delta \Phi_{00} - \bar{\delta} \Phi_{01} + 2 D \Lambda = - \lambda \Psi_{0} + 2(\pi - \alpha) \Psi_{1} + 3 \rho \Psi_{2}\\
    - 2 \kappa \Psi_{3} + (2 \gamma + 2 \bar{\gamma} - \bar{\mu}) \Phi_{00} - 2 (\bar{\tau} + \alpha) \Phi_{01} - 2 \tau \Phi_{10} + 2 \rho \Phi_{11} + \bar{\sigma} \Phi_{02} \ ,
    \end{aligned}
 \end{align}

\begin{align}\label{eq:bianchi6}
    \begin{aligned}
    \Delta \Psi_{2} - \delta \Psi_{3} + D \Phi_{22} - \delta \Phi_{21} + 2 \Delta \Lambda = \sigma \Psi_{4} + 2(\beta - \tau) \Psi_{3} - 3 \mu \Psi_{2} + 2 \nu \Psi_{1}\\
    + (\bar{\rho} - 2 \varepsilon - 2 \bar{\varepsilon}) \Phi_{22} + 2 (\bar{\pi} + \beta) \Psi_{21} + 2 \pi \Phi_{12} - 2 \mu \Phi_{11} - \bar{\lambda} \Phi_{20} \ ,
    \end{aligned}
 \end{align}

 \begin{align}\label{eq:bianchi7}
    \begin{aligned}
    D \Psi_{3} - \bar{\delta} \Psi_{2} - D \Phi_{21} + \delta \Phi_{20} - 2 \bar{\delta} \Lambda = - \kappa \Psi_{4} + 2 (\rho - \varepsilon) \Psi_{3} + 3 \pi \Psi_{2}\\
    - 2 \lambda \Psi_{1} + (2 \bar{\alpha} - 2 \beta - \bar{\pi}) \Phi_{20} - 2 (\bar{\rho} - \varepsilon) \Phi_{21} - 2 \pi \Phi_{11} + 2 \mu \Phi_{10} + \bar{\kappa} \Phi_{22} \ ,
    \end{aligned}
 \end{align}

\begin{align}\label{eq:bianchi8}
    \begin{aligned}
    \Delta \Psi_{1} - \delta \Psi_{2} - \Delta \Phi_{01} + \bar{\delta} \Phi_{02} - 2 \delta \Lambda = \nu \Psi_{0} + 2 (\gamma - \mu) \Psi_{1} - 3 \tau \Psi_{2} + 2 \sigma \Psi_{3}\\
    + (\bar{\tau} - 2 \bar{\beta} + 2 \alpha) \Phi_{02} + (\bar{\mu} - \gamma) \Phi_{01} + 2 \tau \Phi_{11} \ ,
    \end{aligned}
 \end{align}

 \begin{align}
    \begin{aligned}
    D \Phi_{11} & - \delta \Phi_{10} - \bar{\delta} \Phi_{01} + \Delta \Phi_{00} + 3 D \Lambda = (2 \gamma - \mu + 2 \bar{\gamma} - \bar{\mu}) \Phi_{00} \\
   & + (\pi - 2 \alpha - 2 \bar{\tau}) \Phi_{01} + (\bar{\pi} - 2 \bar{\alpha} - 2 \tau) \Phi_{10} + 2 (\rho + \bar{\rho}) \Phi_{11} \\
   & + \bar{\sigma} \Phi_{02} + \sigma \Phi_{20} - \bar{\kappa} \Phi_{12} - \kappa \Phi_{21} \ ,
    \end{aligned}
 \end{align}

\begin{align}
    \begin{aligned}
    D \Phi_{12} & - \delta \Phi_{11} - \bar{\delta} \Phi_{02} + \Delta \Phi_{01} + 3 \delta \Lambda = (-2 \alpha + 2 \bar{\beta} + \pi - \tau) \Phi_{02} \\
    & + (\bar{\rho} + 2 \rho - 2 \bar{\varepsilon}) \Phi_{12} + 2 (\bar{\pi} - \tau) \Phi_{11} + (2 \gamma - 2 \bar{\mu} - \mu) \Phi_{01} \\
    & + \bar{\nu} \Phi_{00} - \bar{\lambda} \Phi_{10} + \sigma \Phi_{21} - \kappa \Phi_{22} \ ,
    \end{aligned}
\end{align}

\begin{align}
    \begin{aligned}
    D \Phi_{22} & - \delta \Phi_{21} - \bar{\delta} \Phi_{12} + \Delta \Phi_{11} + 3 \Delta \Lambda = (\rho + \bar{\rho} - 2 \varepsilon - 2 \bar{\varepsilon}) \Phi_{22} \\
    & + (2 \bar{\beta} + 2 \pi - \bar{\tau}) \Phi_{12} + (2 \beta + 2 \bar{\pi} - \tau) \Phi_{21} - 2 (\mu + \bar{\mu}) \Phi_{11} \\
    & + \nu \Phi_{01} + \bar{\nu} \Phi_{10} - \bar{\lambda} \Phi_{20} - \lambda \Phi_{02} \ .
    \end{aligned}
 \end{align}
 \end{subequations}

\end{widetext}

\end{document}